\documentclass{article}

\usepackage{amsfonts}
\usepackage{amsmath}
\usepackage[margin=0.5 cm]{geometry}

\usepackage{color}
\usepackage{fullpage}

\usepackage{hyperref}
\usepackage{graphicx}
\usepackage{float}

\usepackage[labelsep=period]{caption}

\usepackage{tikz}
\usetikzlibrary{matrix}

\begin{document}

\centerline{\Large Spillover modes in multiplex games: double-edged}
\centerline{\Large  effects on cooperation, and their coevolution}

{\small {\vskip 12pt \centerline{Tommy Khoo$^{1,*}$, Feng Fu$^{1,2}$, \& Scott Pauls$^{1}$}

\begin{center}
$^1$ Department of Mathematics, Dartmouth College, Hanover, NH 03755, USA \\
$^2$ Department of Biomedical Data Science, Geisel School of Medicine, \\ Dartmouth College, Hanover, NH 03755, USA
\end{center}
}}

\begin{center}
[$^*$] tommy.za.khoo@gmail.com
\end{center}

{\bf Abstract:}
In recent years, there has been growing interest in studying games on multiplex networks that account for interactions across linked social contexts. However, little is known about how potential cross-context interference, or spillover, of individual behavioural strategy impact overall cooperation. We consider three plausible spillover modes, quantifying and comparing their effects on the evolution of cooperation. In our model, social interactions take place on two network layers: one represents repeated interactions with close neighbours in a lattice, the other represents one-shot interactions with random individuals across the same population. Spillover can occur during the social learning process with accidental cross-layer strategy transfer, or during social interactions with errors in implementation due to contextual interference. Our analytical results, using extended pair approximation, are in good agreement with extensive simulations. We find double-edged effects of spillover on cooperation: increasing the intensity of spillover can promote cooperation provided cooperation is favoured in one layer, but too much spillover is detrimental. We also discover a bistability phenomenon of cooperation: spillover hinders or promotes cooperation depending on initial frequencies of cooperation in each layer. Furthermore, comparing strategy combinations that emerge in each spillover mode provides a good indication of their co-evolutionary dynamics with cooperation. Our results make testable predictions that inspire future research, and sheds light on human cooperation across social domains and their interference with one another.

\section*{Introduction}

The ubiquity of cooperation in human societies and nature is a puzzling phenomenon \cite{axelrod,TriversQRB71,levin06, IPD1}. At first glance, cooperation seems unlikely: cooperators incur cost in providing benefits to others, while opportunistic individuals can reap rewards without returning the favour \cite{hardin}. Nonetheless, cooperation can arise in structured populations through the mechanism of network reciprocity \cite{5rules,nowak05,santos3}. This basic observation drove deeper investigations into reciprocity in structured populations \cite{Szabo07,OhtsukiNature06,santos2, per08}.

Real world networks are often interdependent, where a small perturbation to one network can trigger a chain of events that results in cataclysmic effects on both networks \cite{cascade}. Taking the importance of such interconnectedness into account has led to a recent boom in the study of multiplex networks \cite{networks3, networks4}. In the same vein, evolutionary games on multiplex networks are attracting increasing attention (we refer readers to \cite{colloquium} for a review).

Various mechanisms have been proposed to associate different evolutionary games taking place on otherwise disjoint networks. One utility function approach incorporates payoffs accumulated across games on different networks into each strategic decision \cite{utility1,utility2}. An alternative approach allows strategic behaviour to be transmitted from one setting to another through peer influence and social learning \cite{neighbors1,neighbors2}. 

In this paper, we draw inspiration from empirical results \cite{spillover2, spillover3, spillover4, spillover5, spillover6}, which suggest that norms and heuristics cultivated during repeated interactions could ``spill over'' to affect decision making in one-shot situations. 

In two experiments \cite{spillover2,spillover5}, when subjects first participate in the iterated prisoner's dilemma (IPD) \cite{IPD}, they observed greater cooperation and prosocial behaviour in subsequent one-shot games. A similar increase in prosocial behaviour follows a repeated public goods game with conditions favourable for cooperation \cite{spillover6}. Relatedly, cooperation levels rose when switching from an IPD with a large continuation probability to one with a small continuation probability \cite{spillover3}, as well as when switching from playing an IPD with a fixed partner to playing an IPD in which every iteration was played with a random partner \cite{spillover4}.

The hypothesized explanation \cite{spillover2,spillover1,rand14} for these phenomena is that repeated interactions foster cooperative heuristics or norms in participants which then affected subsequent one-shot games \cite{IPD1,IPD2}. Although recent endeavours \cite{spillover1,additional1,additional2} studied spillover using an evolutionary framework, systematic exploration of spillover mechanisms through the lens of games on multiplex networks are still lacking.

We fill this gap by modelling spillover as strategy interference between layers on a multiplex network (see Fig. 1A). Instead of the unstructured populations seen in the experiments, our $n$ individuals participate in two layers of interactions with population structure that emulate recurring close proximity and distant one-shot contacts. 

We represent recurring close proximity contacts using a square periodic lattice on the bottom layer where agents play $m$ rounds of the IPD with their neighbours using two possible strategies $C$ or $D$. Here, $C$ refers to tit-for-tat (TFT), where individuals cooperates on the first iteration and play the opponent's previous strategy for future iterations, and $D$ stands for always defecting (ALLD) during every iteration. While other strategies are possible, we focus on these two classic strategies \cite{tft1,tft2} for simplicity. If two players play $C$, they both receive a payoff of $m(b-c)$. If they both play $D$, they both receive $0$. If one plays $C$ and the other plays $D$, the $C$ player gets a payoff of $-c$ while the $D$ player gets $b$.

On the top layer, the same agents, randomly connected to four different agents during every run, play the (one-shot) prisoner's dilemma \cite{PD} as a proxy for distant one-shot contacts. Here, agents choose to cooperate ($C$) or defect ($D$). Two cooperators will both receive a payoff of $b-c$, while two defectors both receives $0$. If one cooperates and the other defects, the cooperator gets a payoff of $-c$ while the defector gets $b$.

We propose three modes for spillover. In the first, which we call neighbour imitation spillover (NIS) mode Fig. 1B, individuals on one layer may imitate the strategy of a neighbour on the opposite layer. In the second, self comparison spillover (SCS) mode Fig. 1C, individuals compare their payoffs between layers and learn from their experience. Finally, for the third context interference spillover (CIS) mode, illustrated in Fig. 1D, individuals may make a temporary mistake and use their strategies from another layer. A parameter $p$ determines the frequency of spillover occurrences in each case, and hence is a proxy for the strength of the spillover effect.

Through these three modes, we amalgamate key ideas from prior work. NIS and SCS encapsulates the notion of individuals making a mistake in learning a potentially suboptimal strategy from a different social setting \cite{neighbors1,neighbors2}, either through their own experience or by interacting with others. CIS captures the idea of individuals making implementation mistakes due to confounding two different social settings \cite{spillover1,additional1,additional2}. We also note that our spillover mechanisms model mistakes occurring between distinct network games, as opposed to random errors modelled by mechanisms such as weak selection and mutation.

Our main finding is that cooperation depends subtly on the strength $p$ and the initial level of cooperation on both the layers - a double-edged effect where different combinations encourage or discourage cooperation. These results expanded upon previous work containing an alternative formulation of NIS \cite{neighbors2}, which used two one-shot games to investigate neighbour imitation and found that there is an intermediate optimal frequency for cooperation in one of the games, in the case of a well-mixed population. We also advance their results for unstructured populations by using pair approximation \cite{pair} to derive analytical solutions that incorporate population structure, and demonstrate the effectiveness of our solutions with extensive simulations. These solutions allow us to study both the macroscopic overall cooperation level, as well as microscopic details regarding strategy combinations.

Finally, we analyse the three spillover modes as they coevolve with cooperation in the presence of mutation. Our findings suggest that transient implementation mistakes (CIS) outperforms mechanisms under which individuals might learn and retain suboptimal strategies (NIS, SCS). On the other hand, when we allow repeated local interactions to play a larger role during spillover conditions become more conducive to cooperation, making it less punishing to make learning mistakes with long lasting impact. This effect allows more deliberate mechanisms that promotes cooperation on both layers, like NIS and SCS, to thrive.

\section*{Results}

We start our exploration by examining how the strength of the spillover effect impacts cooperation. To accomplish this, we produced simulation and pair approximation results for a range of parameter combinations. We present these in Fig. 2, with individual plots for each mode (NIS Fig. 2A, SCS Fig. 2B, and CIS Fig. 2C).

The most striking feature of Fig. 2 is the existence of an optimal value of $p$ which maximizes the average cooperation level in the multiplex network. Average cooperation initially increases with $p$, before reaching the optimal value and subsequently plummeting. This demonstrates a double-edged effect of spillover: a little spillover between the two social settings allows cooperators on the repeated local interaction layer to exert their influence on the distant one-shot contact layer and provides a boost to overall cooperation. On the other hand, a spillover effect that is too strong leads to too much influence by defectors on the distant one-shot contacts layer, and is deleterious to cooperation. We note that conditions that are overly favourable or hostile to cooperation will lead to one layer overwhelming the other and consequently a rapid monotonic rise or decline in cooperation, instead of an intermediate optimal value of $p$ (see Supplementary Figure S1).

Another feature of Fig. 2 is that CIS appears to be more resilient to this double-edge effect than NIS and SCS as cooperation levels for CIS tend to be higher than the other two. This is due to the fact that while implementation mistakes occur under CIS, payoff comparison always occurs on the layer on which the strategies are adopted. Hence, individuals have an easier time learning correct strategies, and implementation mistakes need not have a prolonged impact. On the other hand, under NIS and SCS, individuals directly learn and adopt strategies across layers, resulting in mistakes that have larger, long term repercussions on cooperation.

Next, we examine fine details for several parameter combinations from Fig. 2A-C highlighted with squares. These microscopic details (Fig. 2D-F) show the proportions of individuals for each possible strategy combination for the top and bottom layer. For simplicity, we let $X^a_b$ be the proportion of individuals playing strategy $a$ on the top layer and $b$ on the bottom layer.

Fig. 2D-F demonstrates that for small values of $p$, there is excellent agreement between simulation and our pair approximation results at even the microscopic level (detailed equations are presented in the SI). The most outstanding feature is that context interference spillover mode has much higher $X^d_c$ than the other modes, as seen in Fig. 2F. This happens because, the parameter combinations in Fig. 2 leads to almost all $C$ on the bottom layer and almost all $D$ on the top layer. As we saw above, in CIS, players learn strategies more easily compared to NIS and SCS, so individuals can learn the optimal strategy of playing $D$ on the bottom layer and playing $C$ on the top layer.

On the other hand, SCS (Fig. 2E) stands out as having the highest proportion of individuals who are cooperators on both layers, with NIS (Fig. 2D) coming in a close second. In both cases, individuals are adopting spillover strategies that their neighbours or themselves have been successfully using within the opposite layer. This leads to a higher level of cooperation in the one-shot PD layer due to individuals learning and retaining the suboptimal strategy of playing $C$ on that layer.

Fig. 3 further illuminates the differences in microscopic details between the three spillover modes for a subset of the parameter combinations in Fig. 2. Fig. 3A-C shows SCS having the highest $X^c_c$, while Fig. 3D-F shows CIS having the highest $X^d_c$. We get a clearer view of how NIS differs from the rest, with higher $X^d_d$ and $X^c_d$ (Fig. 3G-L). Fig. S2 and S3 extends Fig. 3 for a range of $p$ from 0 to 0.40.

In our previous results, we have initialised individual strategies $C$ or $D$ on both layers uniformly at random. But how will spillover behave when the initial probability of being a cooperator varies on each layer? We address this question by exhaustively exploring the parameter space using pair approximation as illustrated by Fig. 4 (NIS Fig. 4A, SCS Fig. 4B, CIS Fig. 4C).

Here, we discover a bistability phenomenon. For a fixed frequency of spillover $p$, the parameter space is partitioned into two distinct regions. Depending on the initial proportion of cooperators on each network layer, the spillover effect can either help or hinder cooperation, as shown in more detailed plots by both simulation and pair approximation (Fig. 4D-I). This bistability phenomenon has potential social policy implications: if the proportion of cooperators in one setting can be actively raised to a sufficient level, spillover can promote overall levels of cooperation.

As shown in Fig. 4, the number of initial cooperators on the bottom layer has a larger impact on whether spillover hinders or helps cooperation than the number of initial cooperators on the top layer. An example of this can be seen in Fig. 4A, where at $p = 0.3$, when the number of initial cooperators on the top layer is close to zero, sufficient number of initial cooperators on the bottom layer can still result in a spillover effect that helps cooperation. In contrast, for close to zero initial cooperators on the bottom layer, no amount of initial cooperators on the top layer will result in a beneficial effect of spillover on cooperation.

So far, we have studied these three spillover modes separately. Next, we will compare them when all three modes are present and are potentially competing with each other. We initialise each individual with a spillover mode uniformly at random and allow the modes to coevolve with strategy while at the same time introducing mutation. Fig. 5 shows our results for various $p$.

Each bar in Fig. 5A-B indicates the proportion of individuals with each of the three spillover mode. For all of the choices of $p$ in Fig. 5A, the highest is CIS, followed by SCS and then NIS, with this trend becoming more prominent as $p$ increases. We explain the dominance of CIS for this parameter combination by the crucial role of individuals playing $D$ on the top prisoner's dilemma layer, and $C$ on the bottom iterated prisoner's dilemma layer (Fig. 3). When $p$ is small, the top layer is approximately at full defection, while the bottom layer is approximately at full cooperation. So, playing $D$ on top and $C$ on the bottom offers the highest total payoff on average. As we saw in Fig.s 2F and 3, CIS shows a relative abundance of individuals with this type of mixed strategy, providing a convincing explanation for the supremacy of CIS in the competition.

SCS mode produces a relatively high proportion of individuals cooperating on both layers, which might intuitively be what one would desire in such a social system. However, this tendency results in a loss of individuals playing the optimal combination of strategies ($X^d_c$), leading to a loss of competitiveness when pitted against other spillover modes, under the parameter combination of Fig. 5A. The dominance of CIS appears to be robust when the cost of cooperation $c$ was lowered to $c = 0.30,0.25$ and $0.20$ (see Fig. S4).

However, at $c = 0.20$, this dominance is reduced. Conditions are favourable for cooperation, so we have high equilibrium proportions of cooperators on both layers (Figure S5A). In this situation, $X^c_c$ has a higher payoff than $X^d_c$ and hence, NIS and SCS become more competitive. This increase in competitiveness of the strategy that cooperates on both layers can be overcome by an increase in $p$ (Figure S4A for $p  = 0.05$). However, this does not happen when cooperation on both levels are high enough, and CIS once again loses its advantage (Fig. S4A and S5A, show the value $c = 0.20, p = 0.1$).

In Fig. 5B, we consider an additional parameter $\alpha \in (0,1)$ which governs the relative influence of the two layers. When spillover occurs under NIS and SCS, $\alpha$ is the probability that an individual on the top layer is chosen to possibly learn a strategy from the bottom layer. Under CIS, $\alpha$ is the probability that an individual uses her bottom layer strategy during spillover. We make these definitions so that parameter $\alpha$ consistently refers to how strong an influence the bottom IPD layer has when spillover occurs

We find that higher $\alpha$ promotes cooperation in general (Fig. 5C-D) and this could potentially alter the results of coevolution. As shown in Fig. 5B, it is possible for SCS to be favoured by selection instead of CIS when we set $\alpha = 0.95$. This happens because there is a much higher level of cooperation in both layers when $\alpha$ is high (Fig. 5D and S6B) which allows $X^c_c$ to be more competitive than $X^d_c$. We show comparison of the cooperation level on both layers over time for both the $\alpha = 0.5$ and $\alpha = 0.95$ scenarios described in Fig. S7 and S8.

\section*{Discussion}

Our results generate testable hypotheses that can inspire future research. Several experiments \cite{spillover2,spillover3,spillover4,spillover5} in the literature had participants play the iterated prisoner's dilemma and then switch to various versions of one-shot games. All of these cases reported that the iterated prisoner's dilemma, with conditions that favour cooperation, had a positive effect on cooperation level in the subsequent one-shot game.

Using setups similar to the experiments, one of the spillover mode could be incorporated to test for the existence of an optimal frequency of spillover $p$. For instance, NIS could be implemented experimentally by periodically hiding or revealing the strategies of neighbours on a layer. Similarly, CIS could be implemented by occasionally hiding or mislabelling the layers. At the same time, this can be used to test if spillover modes differ in the proportion of individuals playing each of the four possible top-bottom strategy combinations (Fig. 3), which offers a novel method for comparing and categorizing the myriad of spillover mechanisms that are possible.

Finally, our findings regarding the bistability that arise from varying the initial levels of cooperation on each layer (Fig. 4) could be leveraged to promote cooperation through spillover. As was done in a recent human behaviour experiment \cite{bots}, mixing automated bots with human subjects can lead to the desired cooperation level.

\section*{Materials and methods}

In our model, $n$ individuals are placed on two network layers. The top layer, $T$, is a random regular network of degree four, regenerated with every run, while the bottom layer, $B$, is a two dimensional lattice with periodic boundaries. Individuals play a special version of the prisoner's dilemma game, known as the donation game, with their neighbours on the top layer: they are initially a cooperator $C = [1,0]^T$ or defector $D = [0,1]^T$ with equal probability. The payoff matrix $M_T$ that we use for prisoner's dilemma is,

$$ 
M_T = 
\begin{bmatrix}
b - c & -c \\
b & 0 \\
\end{bmatrix}. 
$$

where $b$ is the benefit of cooperation, while $c$ is the cost of cooperation. On the bottom layer, individuals play the iterated prisoner's dilemma game with their partners, where the game is repeated $m$ times. They initially start with the strategy $C$ or $D$ with equal probability. In this case, $C$ refers to tit-for-tat (TFT), where the individual cooperates on the first iteration and plays the opponent's previous strategy for future iterations. Here, $D$ stands for always defecting (ALLD) during every iteration. The payoff matrix $M_B$ for this is,

$$ 
M_B = 
\begin{bmatrix}
m(b - c) & -c \\
b & 0 \\
\end{bmatrix}. 
$$

Let $s^k_i \in \{ [1,0]^T, [0,1]^T \}$ be the strategy of individual $i$ on layer $k = \{ T,B \}$, and $M_k$ be the payoff matrix of the game on layer $k = \{ T,B \}$. Then, the total payoff of an individual $i$ on layer $k$ is given by,

$$ P^k_i = \sum_{j \in \mathcal{N}^k_i} (s^k_i)^T M_k s^k_j,$$

where $\mathcal{N}^k_i$ is the neighbourhood of $i$ on layer $k$. \\

At each discrete time step, we randomly choose a focal individual. Under NIS and SCS, with probability $1-p$, this individual updates her strategy. Otherwise, with probability $p$, spillover occurs. Under CIS, this individual always updates her strategy using a modified procedure described below. \\

{\bf Strategy updating.}  If individual $i$ chooses to update her strategy, the top layer is chosen as the focal layer with probability $\frac{1}{2}$. Otherwise, the bottom layer is chosen as the focal layer. Next, one of her neighbour $j$ on the focal layer is picked at random. Then, the probability that $i$ copies the strategy of $j$ on the focal layer $k$ is given by the Fermi equation \cite{update1,update2},

$$ F(s^k_j \to s^k_i) = \frac{1}{1 + e^{-\beta(P^k_j - P^k_i)}}, $$

where parameter $\beta$ determines the intensity of selection, and $P^k_i,P^k_j$ are the total payoffs within the layer $k$, of the focal individual $i$ and the neighbour $j$ respectively. \\

{\bf Neighbour Imitation.} If spillover occurs under NIS, we choose the top layer to be the focal layer with probability $\alpha$. Otherwise, with probability $1-\alpha$, we choose the bottom layer. A neighbour on the layer opposite to the focal layer is chosen randomly. Then, the focal individual does payoff comparison and strategy updating on the non-focal layer. However, if she decides to copy this neighbour's strategy, the strategy is instead applied to focal layer, as illustrated by Fig. 1B. \\

{\bf Self Comparison.} If spillover occurs under SCS, we choose the top layer to be the focal layer with probability $\alpha$. Otherwise, with probability $1-\alpha$, we choose the bottom layer. The focal individual then does payoff comparison and strategy updating with herself on the layer opposite to the focal layer, as shown in Fig. 1C. This means that there is a chance for the individual's strategy on the focal layer to be replaced with her strategy on the non-focal layer. However, during SCS, her payoff on the bottom IPD layer is normalized by dividing the payoff matrix throughout by the number of game iterations. The payoff matrix used for SCS is,

$$ \bordermatrix{~ & C & D \cr
                  C & b-c & -\frac{c}{m} \cr
                  D & \frac{b}{m} & 0 \cr}.$$

{\bf Context Interference.} Under CIS, individuals always update their strategy. However, the procedure for doing so is modified. During strategy updating, both the focal individual and the randomly chosen neighbour independently has a probability $p$ of experiencing context interference. If one of them does, she has an independent probability $\alpha$ of using her bottom layer strategy for all parts of the strategy updating procedure. Otherwise, with probability $1-\alpha$, she uses her top layer strategy. These definitions are made so that parameter $\alpha$ consistently refers to how strong an influence the bottom IPD layer has when spillover occurs. This modified strategy updating procedure is shown in Fig. 1D. \\

{\bf Coevolution and mutation.} Fig. 5 was generated by subjecting the three spillover modes to co-evolution and mutation. Individuals initially are assigned a spillover mode at random. We then proceed with two distinct phases: the regular phase and subsequently the coevolution phase. During the regular phase, at each discrete time step, a focal individual is chosen at random, then according to her spillover mode, strategy updating or spillover is carried out as described above. During the coevolution phase, a focal individual and a focal layer are chosen uniformly at random. The focal individual then picks a neighbour on the focal layer at random. Their total payoffs on the multiplex network is then calculated by summing up their total payoffs across both layers. Payoff comparison is then done using the Fermi equation, which gives the probability of the focal individual copying the neighbour's spillover mode. These two phases are run for $10^4$ time steps each, and then repeated in the same order (regular then coevolution phase) until the desired total number of time steps is achieved when summed across all phases. \\

In both the regular and coevolution phases, whenever strategy update is successful, there is a probability $\mu$ of mutation. When mutation occurs, strategy or spillover mode is selected at random, instead of copied. \\

{\bf Analytical solutions.} We derived analytical solutions for each of the spillover mode, taking into account population structure, using extended pair approximation. We refer readers to the SI for the pair approximation equations and their details.

\section*{Data acessibility}
The datasets and code supporting this article have been uploaded as part of the supplementary material.

\section*{Authors' contributions}
T.K., F.F. and S.P. conceived the model. F.F. and T.K. derived the analytical solutions. T.K. wrote the code and analysed the results. F.F. and S.P. supervised the research. All authors reviewed the manuscript.

\section*{Competing interest}
The authors declare that they have no competing interests.

\section*{Acknowledgments}
The work in this paper was supported by the Dartmouth Faculty Start-up Fund to F.F.

\captionsetup[figure]{list=no}

\begin{figure}
\centering
\includegraphics[]{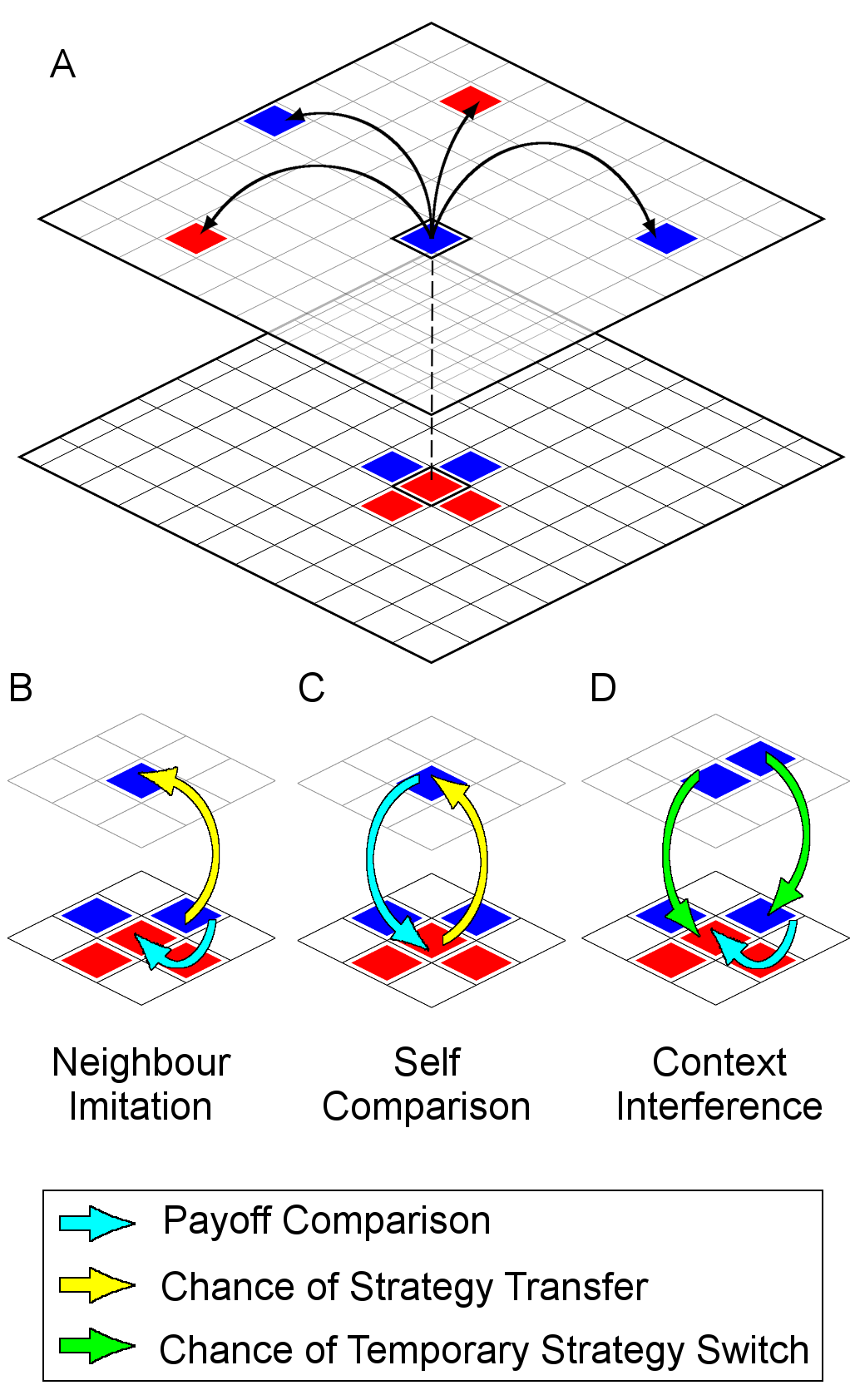}
\caption{{\bf Spillover modes in multiplex games.}
In Fig. 1A, on the bottom layer, individuals play the iterated prisoner's dilemma on a square lattice with periodic boundary. On the top layer, they play the one shot prisoner's dilemma on a random regular network with degree four. Individuals compare payoffs with neighbours within the layer, to decide whether to adopt a neighbour's strategy in that layer. In addition, there is a probability of using one of the three spillover modes instead, to adopt strategies from another layer. For Fig. 1B, neighbour imitation mode, the individual compares payoff with neighbours on the same layer, and applies adopted strategies to the opposite layer. In Fig 1C, self comparison mode, the individual compares her normalized payoffs on each layer, and decides whether to implement her strategy from one layer in another. For Fig 1D, context interference mode, the individual is susceptible to temporary interference from mistaking the context of the interaction, which results in temporarily using strategies from the opposite layer.}
\end{figure}

\begin{figure}
\includegraphics[]{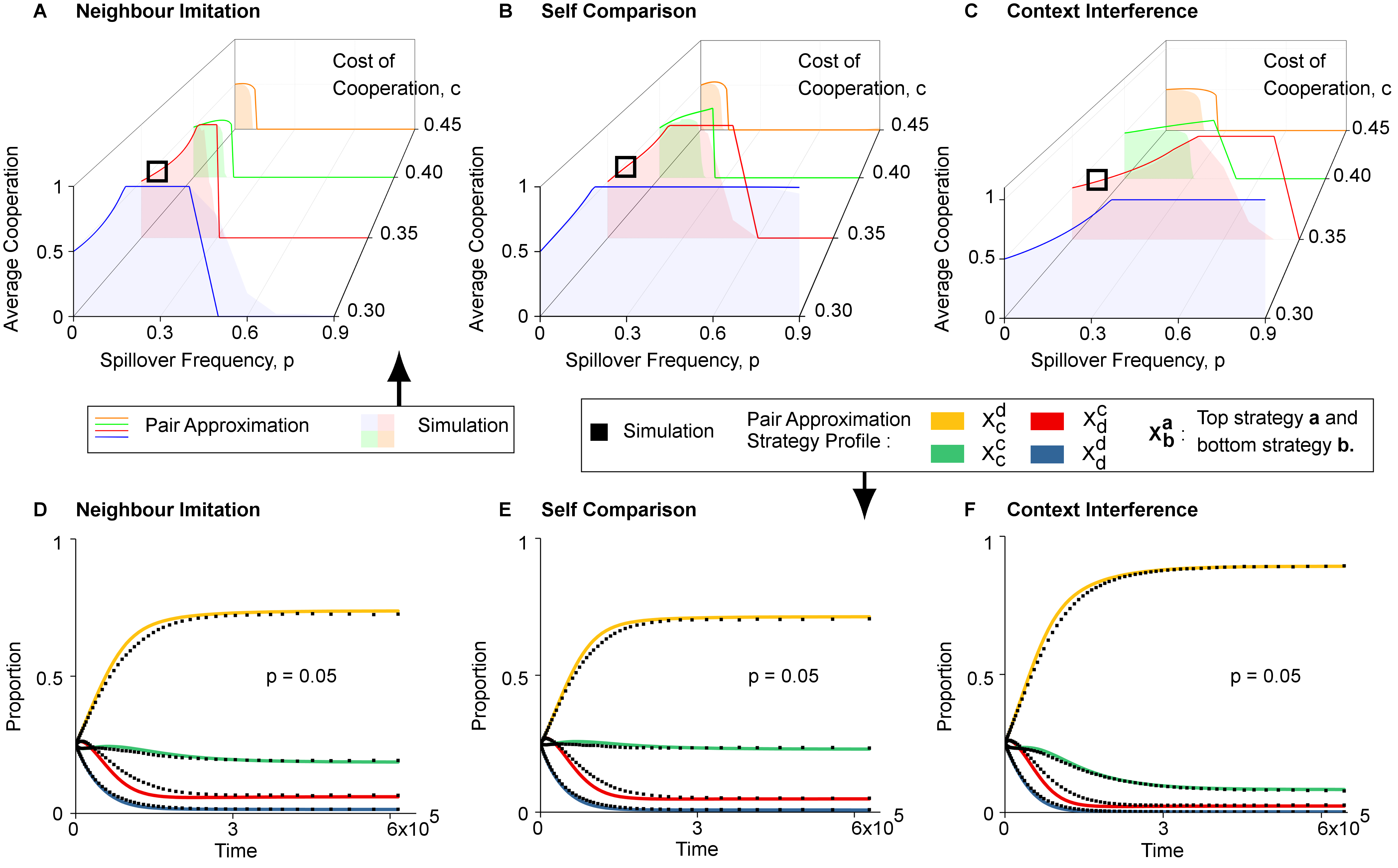}
\caption{{\bf Double-edged effects of spillover.}
Fig. 2A-C shows simulation and pair approximation results for average proportion of cooperators for each spillover mode Fig. 2A neighbour imitation, Fig. 2B self comparison and Fig. 2C context interference. As shown in more details by Fig. 2D-F, analytical and simulation results are in good agreement at low frequencies of spillover $p$. For some parameter combinations, a double-edge effect of spillover on cooperation reveals itself in the form of an initial increase in cooperation with $p$, before a subsequent decline after an optimal $p$. Fig. 2D-F presents microscopic details of strategy profile proportions associated with each spillover mode, taken from Fig. 2A-C at parameter combinations  $(\square)$. Self comparison has the largest proportion of individuals cooperating on both layers (green line), while context interference produces the largest proportion of individuals cooperating on the bottom layer and defecting on the top layer (yellow line). Parameters: $n = 3600$, $m = 4$, $\beta = 0.2$, $b = 1$, $\alpha = 0.5$. Simulations: $6 \times 10^{6}$ time steps, averaged over $100$ runs. $p$ has step size $0.01$ from $p = 0$ to $0.4$, and step size $0.1$ otherwise.}
\end{figure}

\begin{figure}
\includegraphics[scale=0.95]{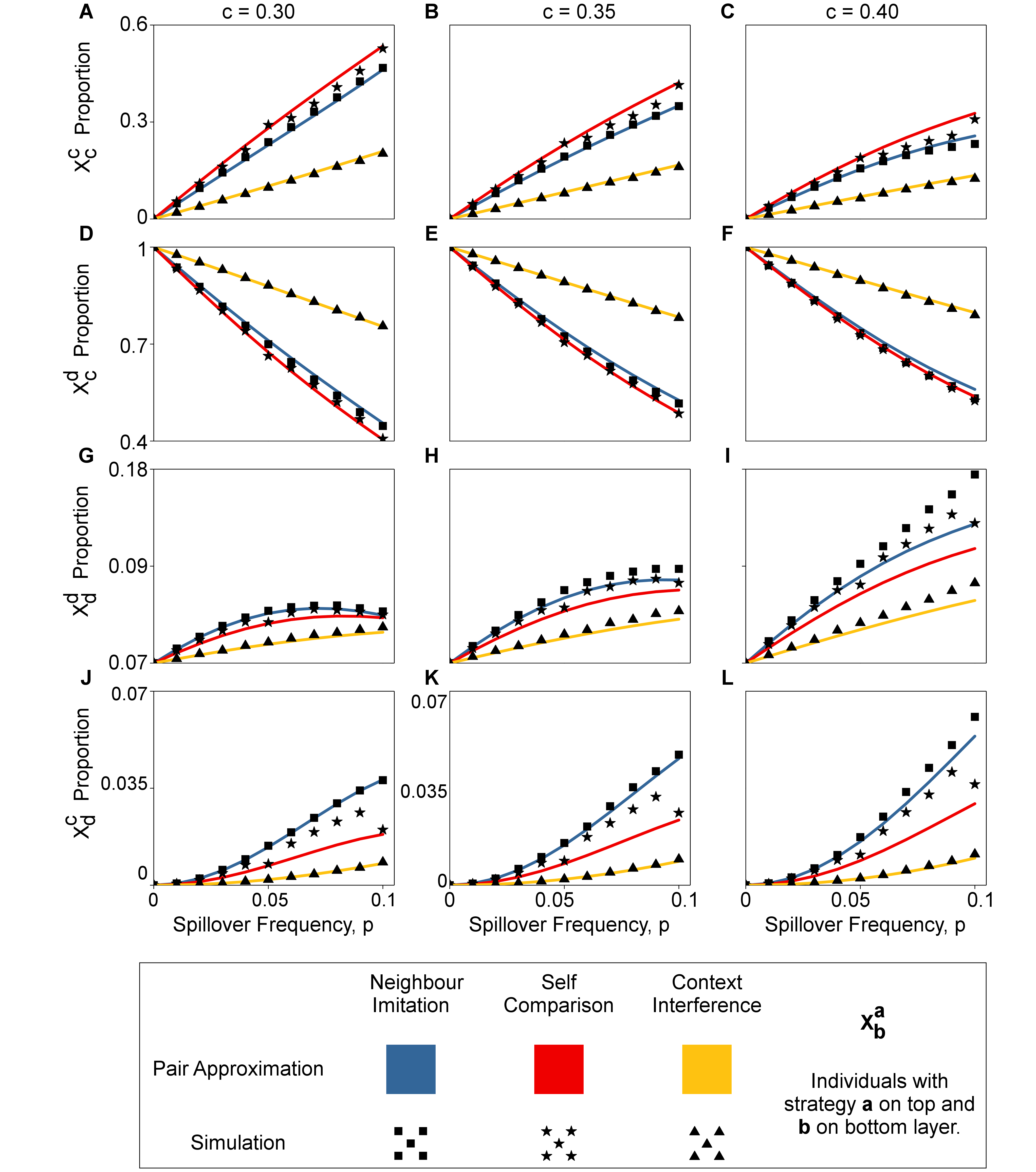}
\caption{{\bf Microscopic characteristics of spillover modes.}
Figure compares proportion of individual top-bottom strategy combinations of each spillover mode for the parameters in Fig. 2, with different scales on the vertical axes. Each spillover mode differs in their microscopic characteristics. In Fig. 3A-C, self comparison has the highest proportion of individuals cooperating on both layers, $X^c_c$. In Fig. 3D-F, context interference has the highest proportion with the payoff maximizing strategy profile $X^d_c$. While in Fig. 3G-L, neighbour imitation has the largest $X^d_d$ and $X^c_d$ proportions. Parameters: $n = 3600$, $m =4$, $b = 1$, $\beta = 0.2$, $\alpha = 0.5$. Simulations: $6 \times 10^{6}$ time steps, averaged over $100$ runs. $c$ is the cost of cooperation.}
\end{figure}

\begin{figure}
\centering
\includegraphics[]{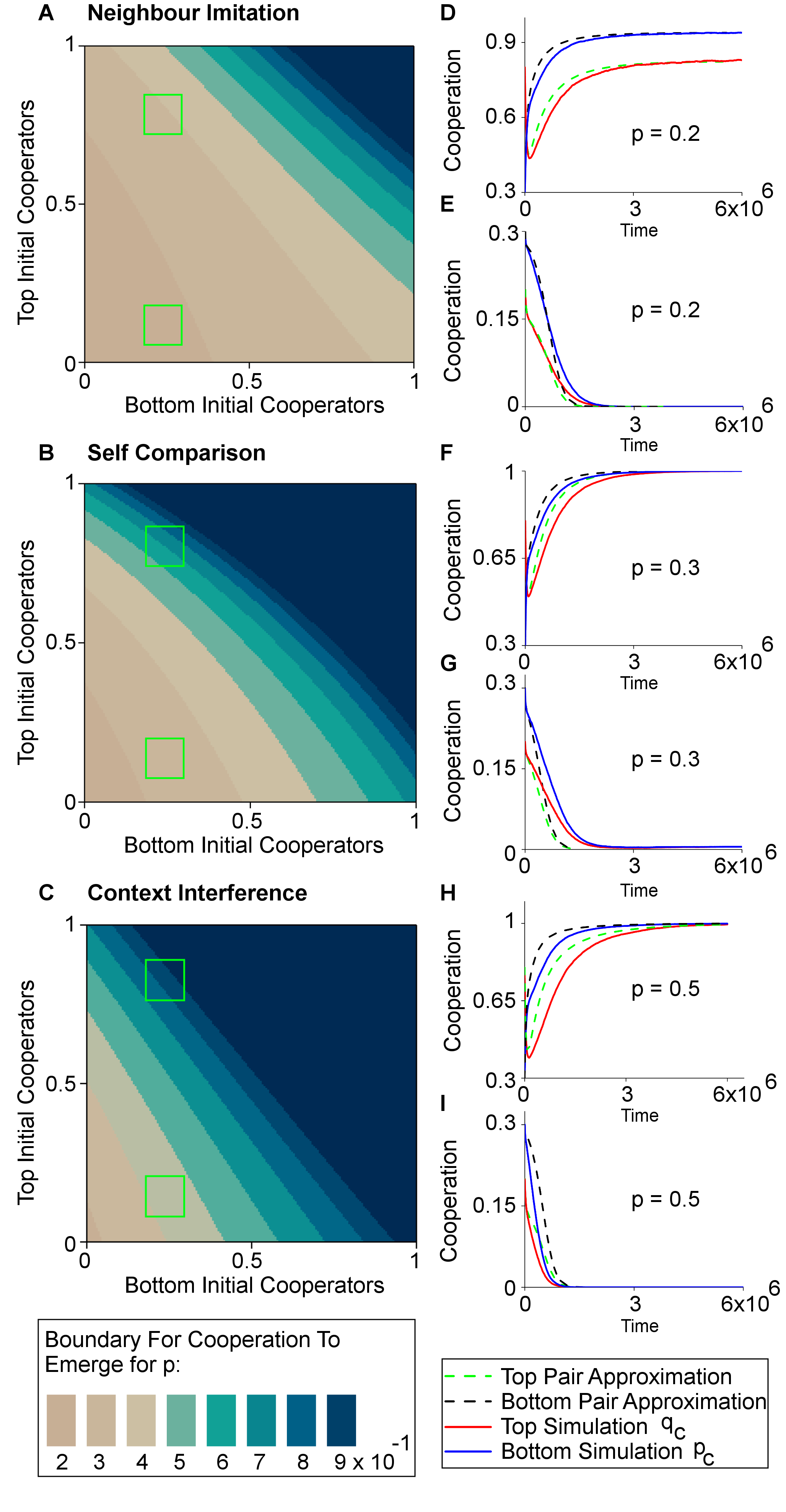}
\caption{{\bf Bistability phenomenon.}
Spillover can hinder or promote cooperation depending on initial conditions. Fig. 4A-C illustrates, for each frequency of spillover $p$, the initial level of cooperators that would result in zero cooperation at equilibrium, under each spillover mode. In all cases, for a fixed $p$, there is a threshold level of initial cooperators on each network layer beyond which the spillover effect switches to working in favor of cooperation. The initial proportion of cooperators required for spillover effect to give rise to cooperation becomes more demanding as $p$ increases. Context interference is shown to be most resilient to this, followed by self comparison, while neighbour imitation is the most vulnerable. Fig. 4D-I demonstrates the bistability phenomenon for selected parameter combinations $(\square)$ with both simulation and pair approximation. Parameters: $n = 3600$, $m = 4$, $c = 0.35$, $b = 1$, $\beta = 0.2$, $\alpha = 0.5$. Data for Fig. 4A-C starts at $0.05$ and ends at $0.95$ initial cooperators for both layers. Simulations in Fig. 4D-I: $6 \times 10^6$ time steps, averaged over $200$ runs.}
\end{figure}

\begin{figure}
\includegraphics[]{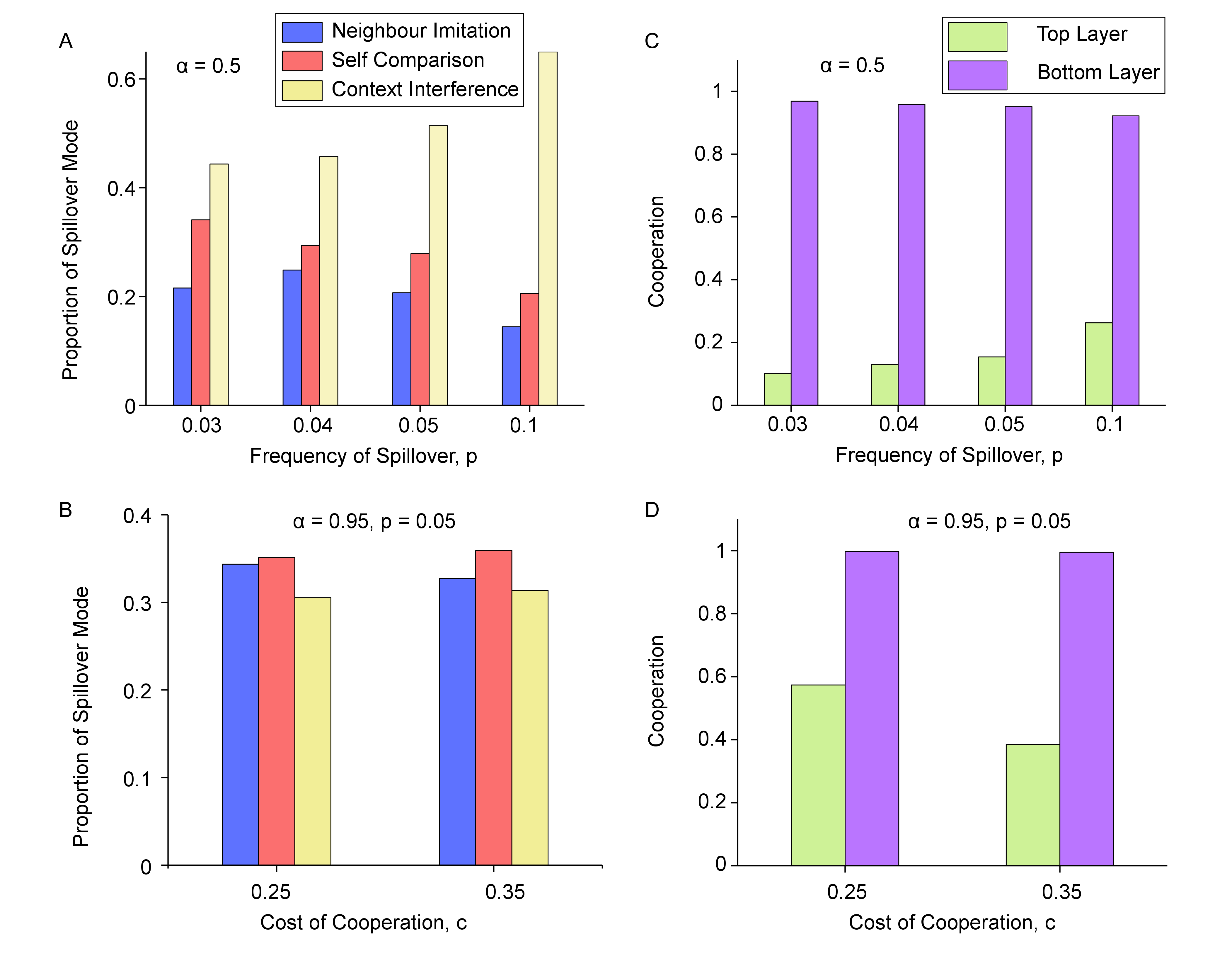}
\caption{{\bf Coevolution of spillover modes and cooperation.}
Fig. 5 shows the proportion of each spillover mode, for low values of $p$, when spillover modes coevolve with strategy in the presence of mutation. Self comparison (red) is second place in Fig. 5A despite having the most individuals who cooperate on both layers. Context interference (yellow) has the largest proportion due to having the most individuals playing the payoff maximizing combination of C on the bottom layer and D on the top layer. Neighbour imitation (blue) accounts for the least proportion of the population. Fig. 5B shows that when $\alpha = 0.95$, self comparison spillover mode can have the highest proportion instead. A high $\alpha$ allows repeated interactions in the bottom layer to have a large influence and creates very favourable conditions for cooperation, as shown by Fig. 5C-D. This allows for individuals cooperating on both layers, which is plentiful under SCS, to become competitive during coevolution. Parameters: $n = 400$, $m = 4$, $c = 0.35$, $\beta = 0.2$, $b = 1$. Mutation rate $\mu = 10^{-4}$, $7.2 \times 10^{9}$ total time steps, combined from at most $7$ runs for Fig. 5A and 5C. $\mu = 10^{-3}$, $1.4 \times 10^{9}$ total time steps, combined from $6$ runs for Fig. 5B and 5D.}
\end{figure}

\captionsetup[figure]{list=yes}

\newpage

\begin{center}
\section*{Spillover modes in multiplex games: double-edged effects on cooperation, and their coevolution}
\subsection*{Tommy Khoo, Feng Fu, Scott Pauls}
\end{center}

\section*{Supplementary Information}

\listoffigures
\tableofcontents

\newpage

\renewcommand{\figurename}{{\bf Figure S}}
\setcounter{figure}{0}
\begin{figure}[h]
\centering
\includegraphics[scale=0.4]{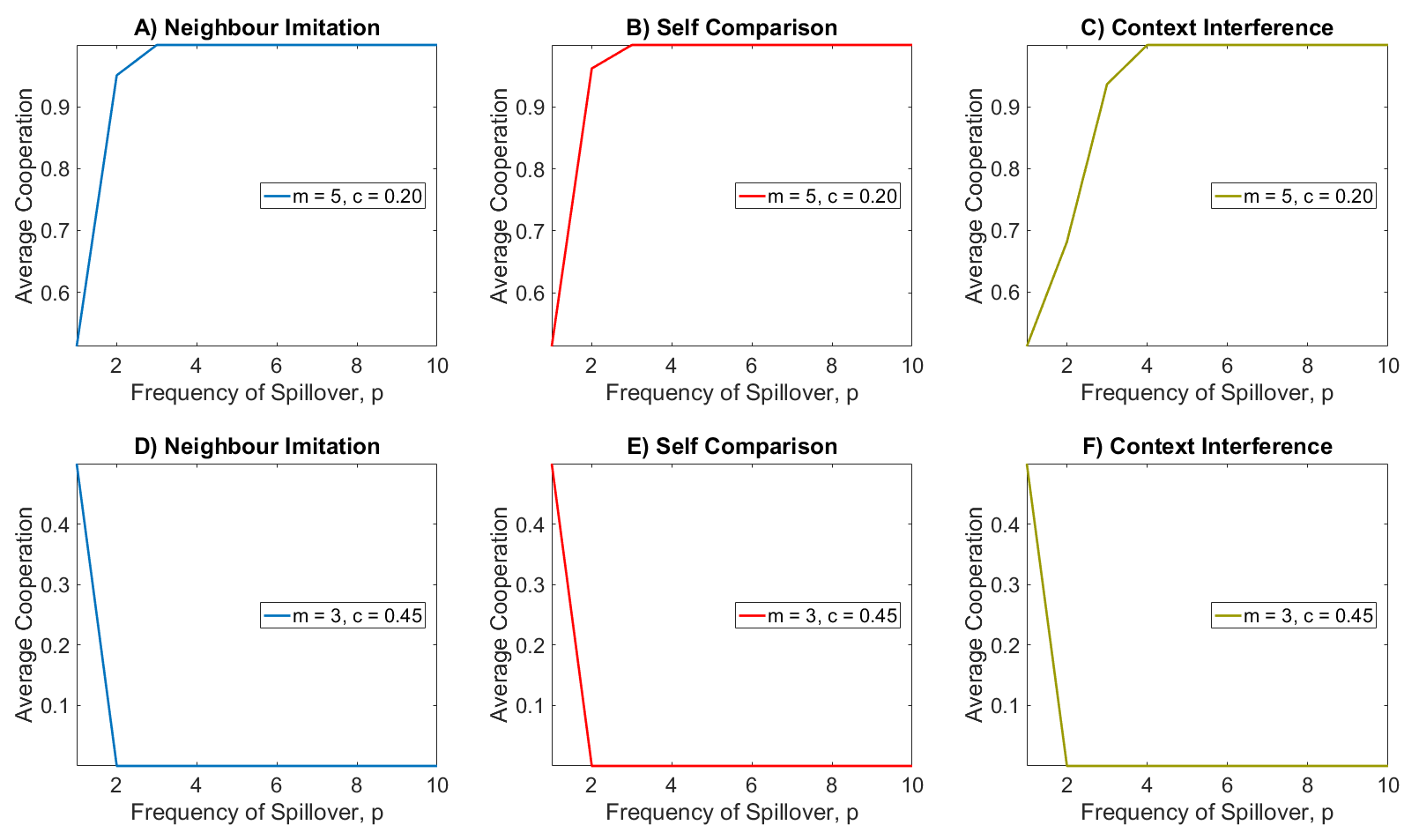}
\caption[Cooperation under highly beneficial or detrimental conditions.]{ {\bf Cooperation under highly beneficial or detrimental conditions.} Figure indicates that when conditions are highly beneficial or detrimental to cooperation, average cooperation level in the multiplex network rapidly rise to one or decline to zero monotonically. As illustrated by Fig. 2 in the main text, an optimal $p$ arise for conditions in between these extremes. Parameters: $\beta = 0.2$, $b = 1$, $\alpha = 0.5$. Results generated using pair approximation.}
\end{figure}

\begin{figure}
\centering
\includegraphics[scale=1]{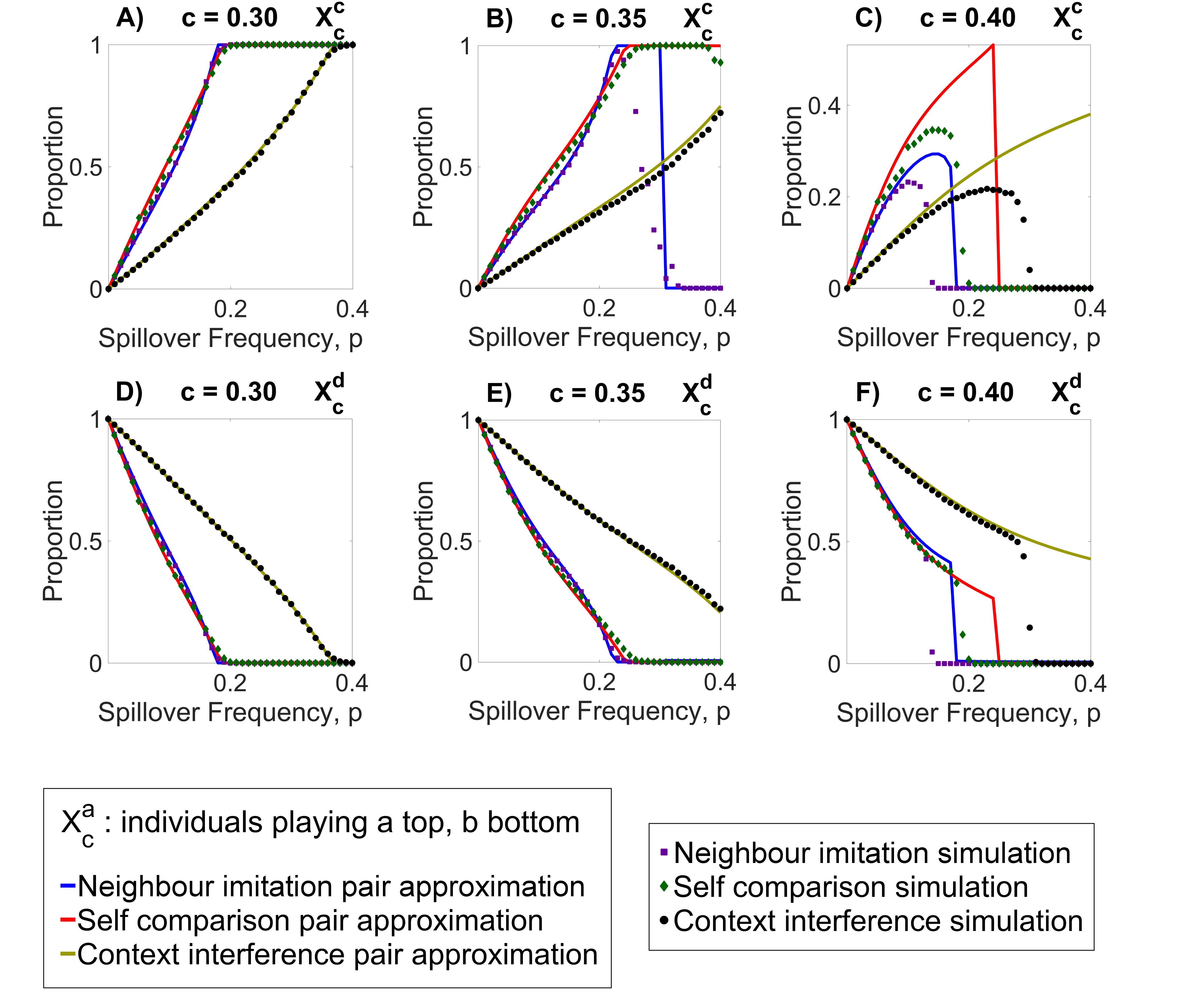}
\caption[Extended Fig. 3 for $X^c_c$ and $X^d_c$.]{ {\bf Extended Fig. 3 for $X^c_c$ and $X^d_c$.} Figure shows an extended version of Fig. 3 in the main text for the proportion of individuals with strategy profile $X^c_c$ and $X^d_c$ respectively, for each spillover mode. Parameters: $n = 3600$, $b = 1$, $\beta = 0.2$, $\alpha = 0.5$. Parameter $p$ range from $p = 0$ to $p = 0.40$ in steps of $0.01$. $c$ is the cost of cooperation. Simulations: $6 \times 10^6$ time steps, averaged over $100$ runs. }
\end{figure}

\begin{figure}
\centering
\includegraphics[scale=1]{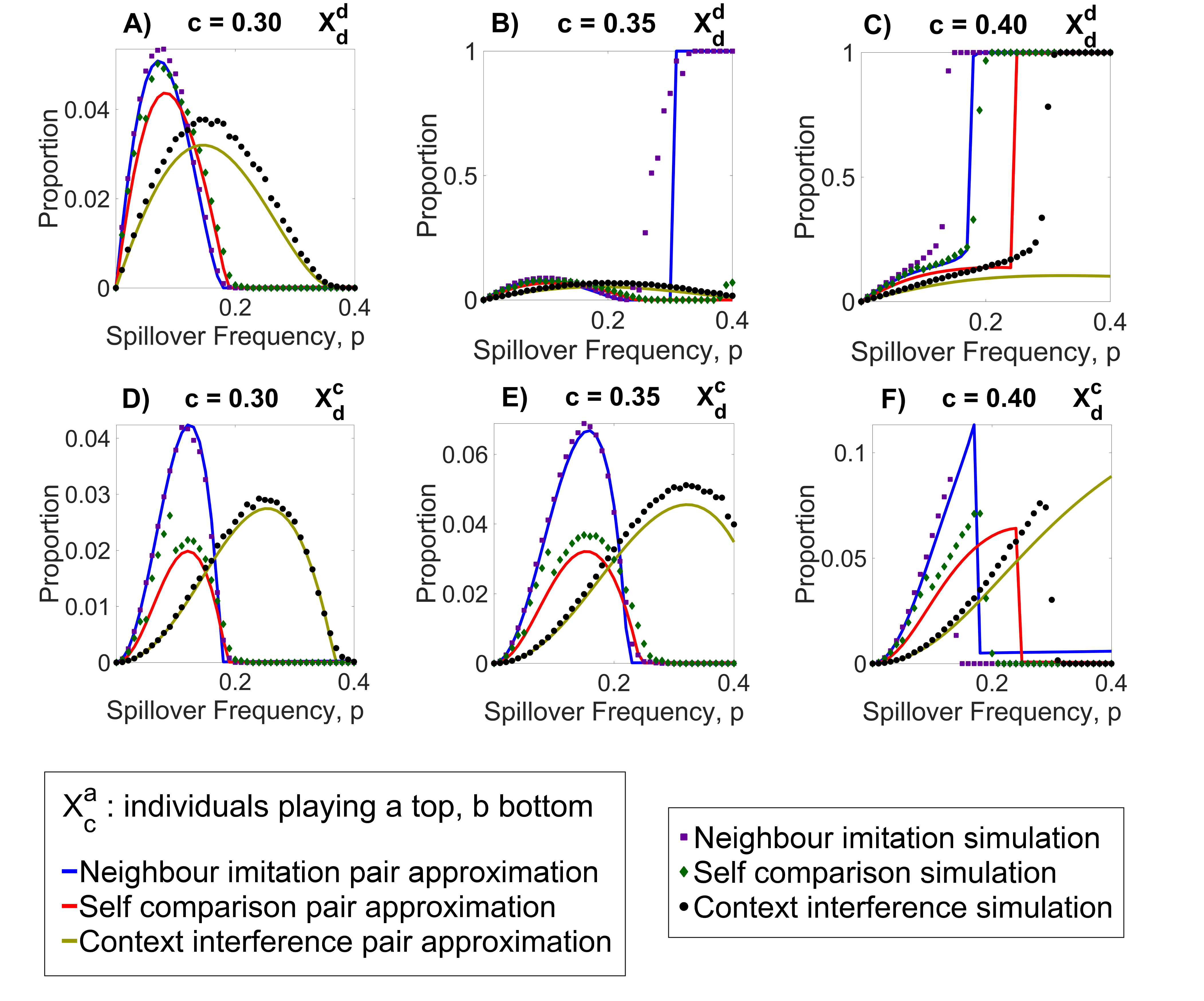}
\caption[Extended Fig. 3 for $X^d_d$ and $X^c_d$.]{ {\bf Extended Fig. 3 for $X^d_d$ and $X^c_d$.} Figure shows an extended version of Fig. 3 in the main text for the proportion of individuals with strategy profile $X^d_d$ and $X^c_d$ respectively, for each spillover mode. Parameters: $n = 3600$, $b = 1$, $\beta = 0.2$, $\alpha = 0.5$. Parameter $p$ range from $p = 0$ to $p = 0.40$ in steps of $0.01$. $c$ is the cost of cooperation. Simulations: $6 \times 10^6$ time steps, averaged over $100$ runs.}
\end{figure}

\begin{figure}
\centering
\includegraphics[scale=1]{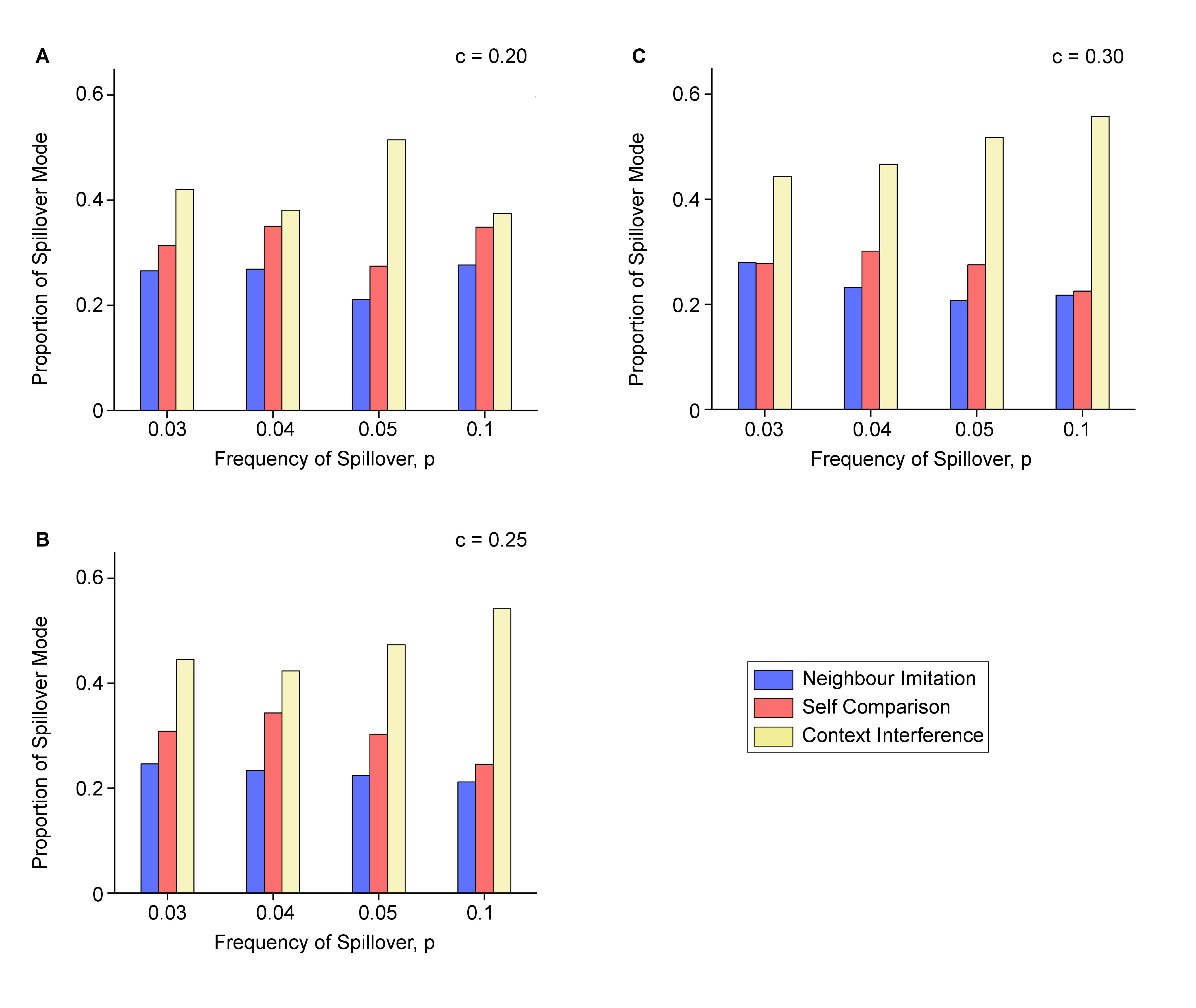}
\caption[Coevolution of spillover modes for $c = 0.30,0.25,0.20$.]{ {\bf Coevolution of spillover modes for $c = 0.30,0.25,0.20$.} Proportion of spillover modes under coevolution and mutation are shown here for parameters different from Fig. 5A in the main text. For cost of cooperation $c = 0.30$, $p = 0.03$, we see that neighbour imitation and self comparison are eventually matched, but the competitive advantage of self comparison is exaggerated with a higher $p = 0.04$. At $c = 0.20$, when conditions are very favourable for cooperation, context interference starts to lose its lead. However, if support for cooperation is not strong enough, context interference can still maintain its lead with a sufficiently high $p$, as shown by the $c = 0.20$, $p = 0.05$ case.  Fig. S5 contain more details explaining these phenomena. Parameters: $n = 400$, $\beta = 0.2$, $b = 1$, $\alpha = 0.5$, mutation rate $\mu = 10^{-4}$. Simulation results consists of at least $4.7 \times 10^9$ total time steps, combined over $6$ runs.}
\end{figure}

\begin{figure}
\centering
\includegraphics[scale=1]{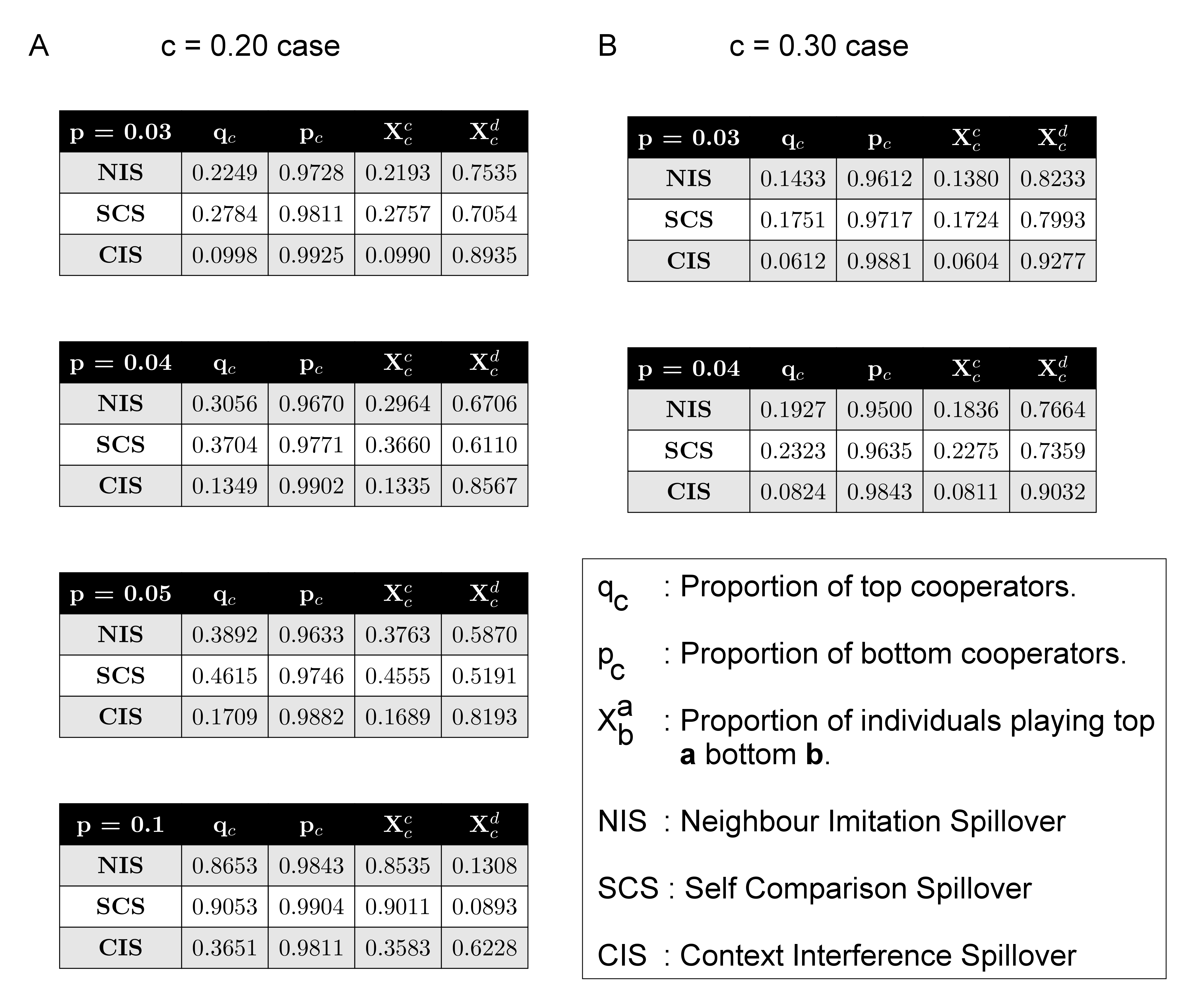}
\caption[Fine details for spillover modes at $c = 0.20$ and $0.30$.]{ {\bf Fine details for spillover modes at $c = 0.20$ and $0.30$. } Figure shows, for each spillover mode, the proportion of cooperators $q_c, p_c$ on each network layer, as well as the strategy profiles $X^c_c, X^d_c$. Fig. S5A explains what we see with the $c = 0.20$ case in Fig. S4A. As $p$ is increased from $0.03$ to $0.04$, we see a rise in cooperation in both network layers, leading to an increase in the competitiveness of $X^c_c$ and the resulting decline in context interference spillover mode. This effect can be suppressed with a higher $p = 0.05$ while cooperation level remains similar. However, as cooperation levels increase dramatically at $p = 0.1$, context interference once again loses its advantage. Parameters: $n = 400$, $m = 4$, $b = 1$, $\beta = 0.2$, $\alpha = 0.5$. Results generated by pair approximation.}
\end{figure}

\begin{figure}
\centering
\includegraphics[scale=1]{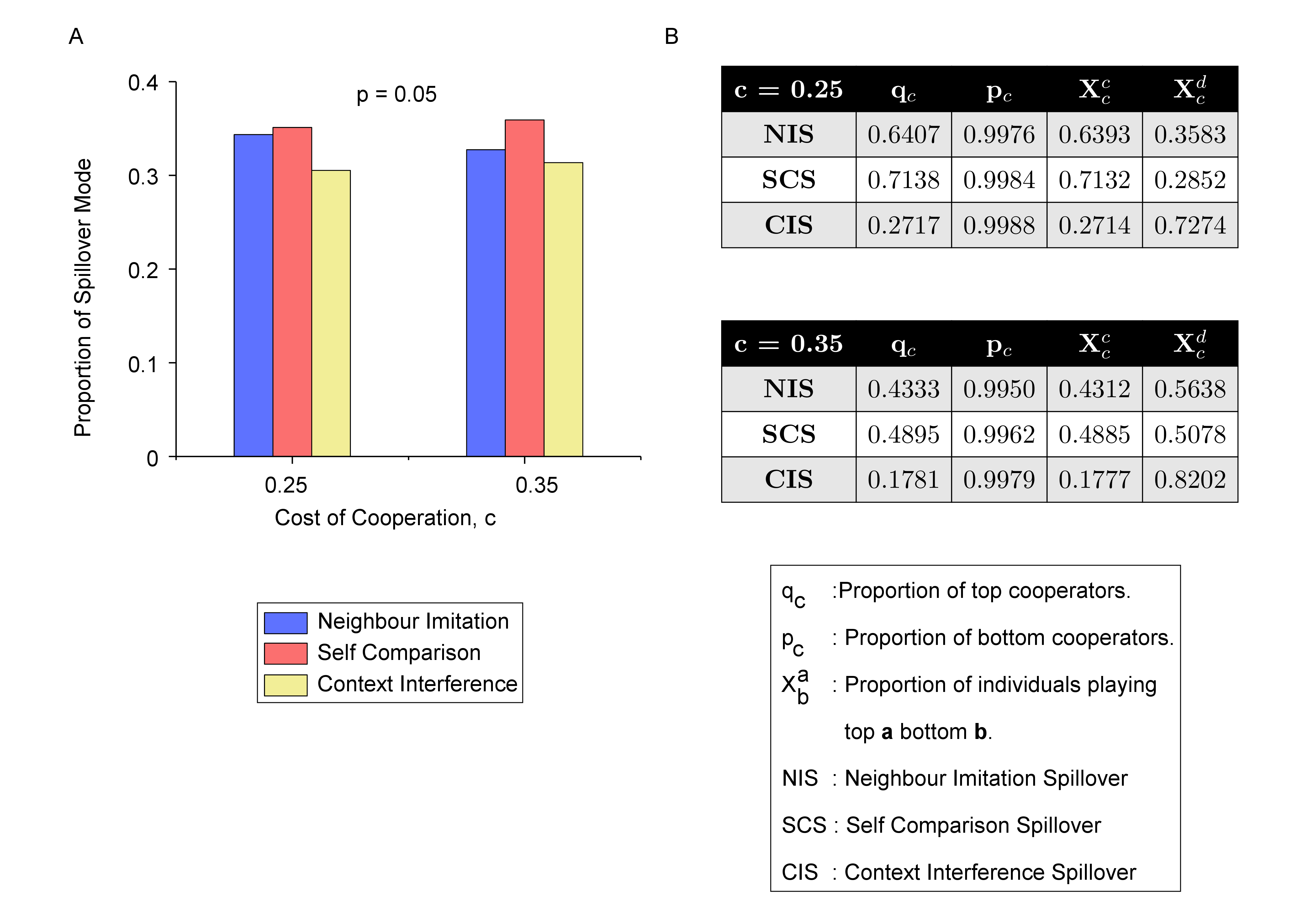}
\caption[Coevolution of spillover modes for $\alpha = 0.95$.]{ {\bf Coevolution of spillover modes for $\alpha = 0.95$.} Figure shows how a high $\alpha = 0.95$ can change which spillover mode takes the lead in the coevolution and mutation case. Parameter $\alpha$ is the probability, during spillover, of selecting the bottom repeated local interaction layer to influence the top one-shot layer. When repeated local interactions have a much higher influence, conditions are much more favourable for cooperation on both network layers. The strategy profile $X^c_c$ can become more competitive when compared to $X^d_c$. Parameters: $n = 400$. $\alpha = 0.95$, $m = 4, b = 1$, $\beta = 0.2$, mutation rate $\mu = 10^{-3}$. Simulation in Fig. S6A has at least $1.4 \times 10^9$ time steps, combined over $6$ runs. Fig. S6B produced using pair approximation. }
\end{figure}

\begin{figure}
\centering
\includegraphics[scale=1]{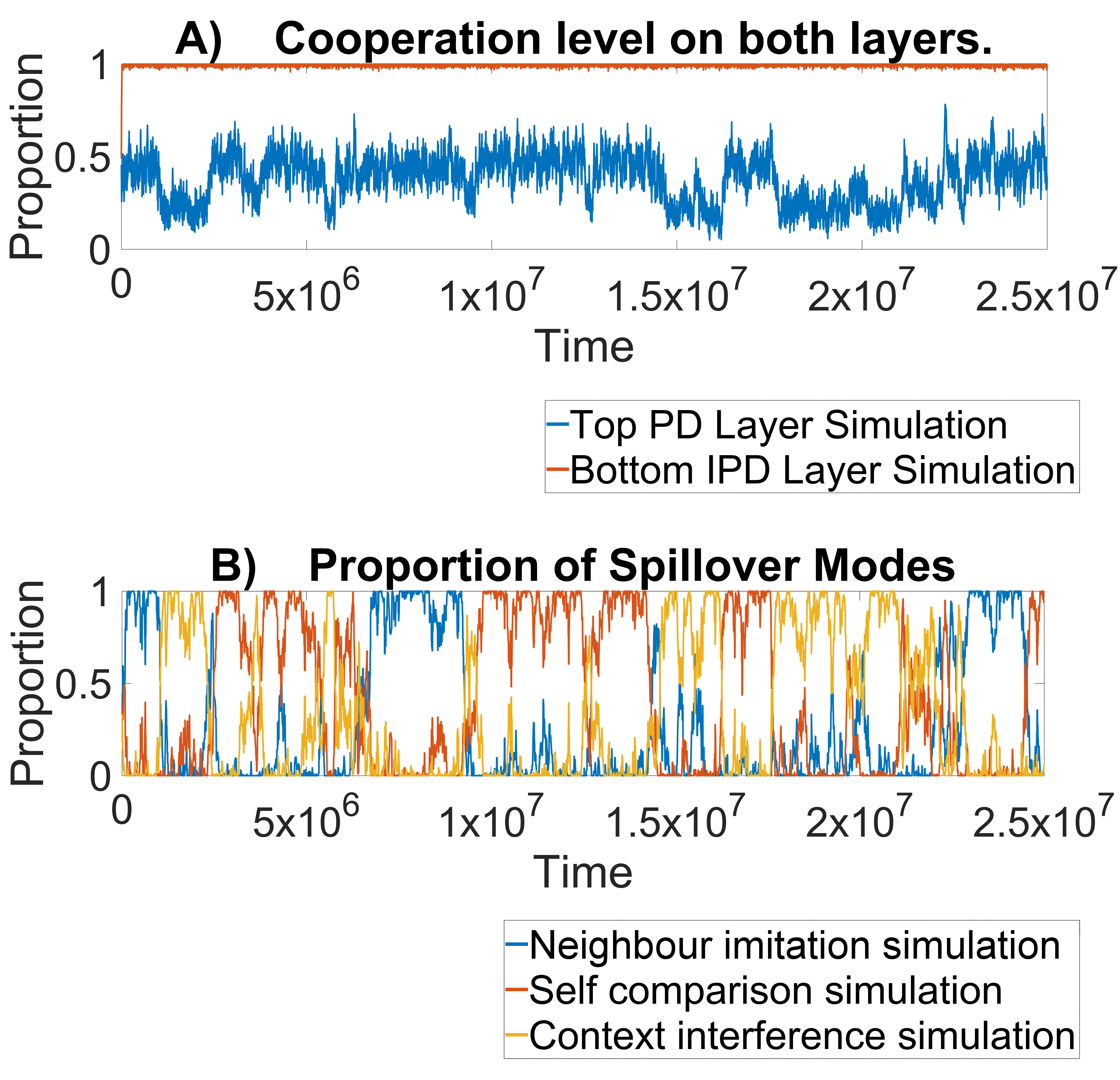}
\caption[Time evolution of cooperation and spillover modes for $\alpha = 0.95, c = 0.35$ case.]{ {\bf Time evolution of cooperation and spillover modes for $\alpha = 0.95, c = 0.35$ case.} Figure demonstrates how neighbour imitation (NIS) and self comparison (SCS) produces a higher level of cooperation in the multiplex when they are dominant, and how context interference (CIS) results in a lower level of cooperation when it is dominant. This lead to NIS and SCS becoming much harder to invade by CIS as $X^c_c$ now has higher competitiveness. Parameters: $n = 400$, $m = 4$, $b = 1$, $\beta = 0.2$, $p = 0.05$, $\alpha = 0.95$, $c = 0.35$. Mutation rate $\mu = 10^{-3}$. Simulation data taken from the first $2.5 \times 10^7$ time steps of a single run. }
\end{figure}

\begin{figure}
\centering
\includegraphics[scale=1]{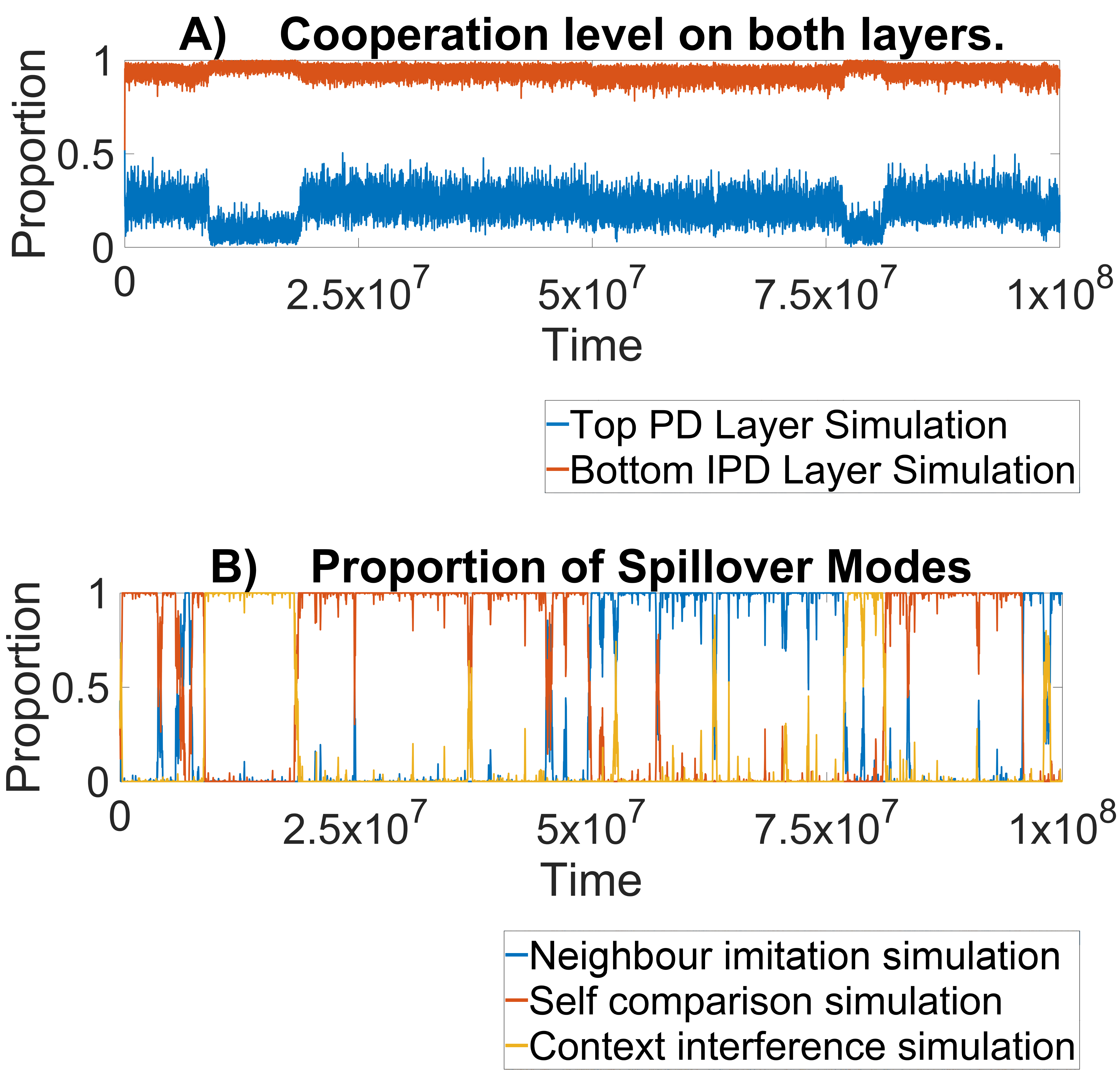}
\caption[Time evolution of cooperation and spillover modes for $\alpha = 0.50, c = 0.35$ case.]{ {\bf Time evolution of cooperation and spillover modes for $\alpha = 0.50, c = 0.35$ case.} Figure illustrates results for a parameter combination from Fig. 5 in the main text. Here, we see that cooperation levels are at a lower level than in Fig. S7. Thus, $X^d_c$ is much more competitive than $X^c_c$, leading to the dominance of context interference spillover mode. Parameters: $n = 400$, $m = 4$, $b = 1$, $\beta = 0.2$, $p = 0.05$, $\alpha = 0.95$, $c = 0.35$. Mutation rate $\mu = 10^{-4}$. Simulation data taken from the first $1 \times 10^8$ time steps of a single run.}
\end{figure}

\newpage

\section{ Datasets and Code }

Datasets and code supporting this article:  Figshare doi:10.6084/m9.figshare.5480548

\section{ Pair Approximation }

We will use pair approximation to derive analytic solutions to each of the three spillover modes. Our analytic solution for each mode consists of differential equations for $\dot{q_c}, \dot{p_c}, \dot{q}_{cc}, \dot{p}_{cc}, \dot{x}^c_c$ and $\dot{x}^d_c$. We start with the notation and definitions common to all three spillover modes.

\subsubsection{ Inter-layer Notation} 

Let $x^a_b$ be the fraction of individuals with strategies $a$ on the top prisoner's dilemma layer, and $b$ on the bottom iterated prisoner's dilemma layer. Letting $c$ be cooperate and $d$ defect, we have the four cases: $x^c_c, x^d_c, x^c_d$ and $x^d_d$, as well as $x^c_c + x^d_c + x^c_d + x^d_d = 1$.

\subsubsection{ Top Layer Notation}

Let $q_c$ and $q_d$ be the top layer fraction of $C$ and $D$ respectively. We must have $q_c + q_d = 1$. Let $q_{ab}$ be the top layer fraction of connected pairs with strategies $a$ and $b$. We have four cases $q_{cc},q_{cd},q_{dc}$ and $q_{dd}$, as well as $q_{cc} + q_{cd} + q_{dc} + q_{dd} = 1$. Let the conditional probabilities be $q_{c|c} = \frac{q_{cc}}{q_c}$, $q_{d|c} = \frac{q_{dc}}{q_c}$, $q_{c|d} = \frac{q_{cd}}{q_d}$, and $q_{d|d} = \frac{q_{dd}}{q_d}$. We must have $q_{c|c} + q_{d|c} = 1$, and $q_{d|d} + q_{c|d} = 1$.

\subsubsection{ Bottom Layer Notation}

Let $p_c$ and $p_d$ be the bottom layer fraction of $C$ and $D$ respectively. We must have $p_c + p_d = 1$. Let $p_{ab}$ be the bottom layer fraction of connected pairs with strategies $a$ and $b$. We have four cases $p_{cc},p_{cd},p_{dc}$ and $p_{dd}$, as well as $p_{cc} + p_{cd} + p_{dc} + p_{dd} = 1$. Let the conditional probabilities be $p_{c|c} = \frac{p_{cc}}{p_c}$, $p_{d|c} = \frac{p_{dc}}{p_c}$, $p_{c|d} = \frac{p_{cd}}{p_d}$, and $p_{d|d} = \frac{p_{dd}}{p_d}$. We must have $p_{c|c} + p_{d|c} = 1$, and $p_{d|d} + p_{c|d} = 1$.

\subsubsection{Top Layer Transition Probabilities Notation and Definitions}

The probability of strategy $b$ replacing strategy $a$ is given by the Fermi equation, 

$$ F(\pi_b,\pi_a) = \frac{1}{1+e^{-\beta(\pi_b - \pi_a)}},$$

where $\pi_j$ is the average payoff from using strategy $j$. Let $T$ be the top layer and $B$ the bottom layer. Let the payoff matrix for layer $i \in \{T,B\}$ be,

$$ \begin{pmatrix} 
R_i & S_i \\
T_i & P_i
\end{pmatrix}.$$ 

Let $k_i$ be the average degree of individuals in layer $i \in \{T,B\}$. Finally, let $W_{b \to a}$ be the probability that, if two individuals on the top layer are chosen for strategy updating, the event ``strategy $b$ replaces $a$" occurs. Then, we have

$$W_{c \to c} = W_{d \to d} = \frac{1}{2}.$$

For $W_{c \to d}$ and $W_{d \to c}$, we need to consider the average payoff of a pair of connected individuals with different strategies. So, we have

$$\pi_c = (k_T - 1) q_{c|c} \cdot R_T + (1+(k_{T}-1)q_{d|c}) \cdot S_T,$$

$$\pi_d = (1+(k_{T}-1)q_{c|d}) \cdot T_T + (k_{T}-1)q_{d|d} \cdot P_T,$$

$$W_{c \to d} = F(\pi_c,\pi_d),$$

$$W_{d \to c} = F(\pi_d,\pi_c).$$

\subsubsection{Bottom Layer Transition Probabilities Notation and Definitions}

Let a pair of connected individuals with strategies $A$ and $B$ be an $AB$ pair. Consider a $CC$ pair in the bottom layer, which is a two dimensional lattice with periodic boundaries. We call the individual that is selected for strategy updating the focal individual, while the randomly chosen neighbour the non-focal individual. Let $x,y,z$ be the strategies of the three other neighbours of the focal individual. Let $u,v,w$ be the strategies of the three other neighbours of the non-focal individual. \\

Let $n(x,y,z)$ be the number of strategy $C$ among the strategies $x,y,z$. Then, the average payoff of the focal individual is now

$$\pi_c = (n(x,y,z)+1) \cdot R_B + (3 - n(x,y,z)) \cdot S_B.$$

The average payoff of the non-focal individual is,

$$\pi'_c = (n(u,v,w)+1) \cdot R_B + (3 - n(u,v,w)) \cdot S_B.$$

Let $\phi_{b \to a}$ be the probability that, if the focal individual is using strategy $a$ and the non-focal individual is using strategy $b$, strategy $b$ is sucessful at replacing strategy $a$. Putting it all together, the probability that a focal $C$ copies a neighbouring non-focal $C$ in the bottom layer is,

$$\phi_{c \to c} = \frac{p_{cx}p_{cy}p_{cz}p_{cu}p_{cv}p_{cw}}{p^{3}_{c}p^{3}_{c}} F(\pi'_{c},\pi_c), $$

we used the fact that $p_{x|c}p_{y|c}p_{z|c} = \frac{p_{cx}p_{cy}p_{cz}p_{cu}p_{cv}p_{cw}}{p^{3}_{c}p^{3}_{c}}$.  \\

For a $DD$ pair in the bottom layer, the average payoff of the focal node and non-focal node are respectively,

$$\pi_d = n(x,y,z) \cdot T_B + (1 +(3 - n(x,y,z))) \cdot P_B,$$

$$\pi'_d = n(u,v,w) \cdot T_B + (1 +(3 - n(u,v,w))) \cdot P_B.$$

Then, the probability that a focal $D$ copies a neighbouring non-focal $D$ in the bottom layer is,

$$\phi_{d \to d} = \frac{p_{dx}p_{dy}p_{dz}p_{du}p_{dv}p_{dw}}{p^{3}_{d}p^{3}_{d}} F(\pi^{'}_{d},\pi_d). $$

Finally, we consider $CD$ pairs. The average payoffs are,

$$\pi_c = n(x,y,z) \cdot R_B + ( 1 + (3 - n(x,y,z)) ) \cdot S_B,$$

$$\pi_d = (1+n(u,v,w)) \cdot T_B + (3 - n(u,v,w)) \cdot P_B.$$

And then the probability that a focal $C$ copies a neighbouring non-focal $D$ in the bottom layer is,

$$\phi_{d \to c} = \frac{p_{cx}p_{cy}p_{cz}p_{du}p_{dv}p_{dw}}{p^{3}_{c}p^{3}_{d}} F(\pi_d,\pi_c). $$

Also, for $DC$ pairs. The average payoffs are,

$$\pi_d = (1+n(x,y,z)) \cdot T_B + (3 - n(x,y,z)) \cdot P_B.$$

$$\pi_c = n(u,v,w) \cdot R_B + ( 1 + (3 - n(u,v,w)) ) \cdot S_B,$$

The probability that a focal $D$ copies a neighbouring non-focal $C$ in the bottom layer is,

$$\phi_{c \to d} = \frac{p_{dx}p_{dy}p_{dz}p_{cu}p_{cv}p_{cw}}{p^{3}_{d}p^{3}_{c}} F(\pi_c,\pi_d). $$

\subsubsection{Notation for Probabilities of Specific Events and Configurations}

In each of the three spillover modes, there are certain configurations of individuals and strategies that can lead to a change in the system. Let $x,y,z$ be strategies. Let the focal node have top strategy $x$ and bottom strategy $z$. Let the non-focal neighbour node have top strategy $y$ and bottom strategy $w$. Then, we represent the event that the top strategy $x$ is replaced with strategy $y$ by,

$$\begin{tikzpicture}[baseline=3ex,scale=1]

\node[draw=none] at (0,1) (a) {x};
\node[draw=none] at (1,1) (b) {y};
\node[draw=none] at (3,1) () {(top layer)};

\node[draw=none] at (0,0) (c) {z};
\node[draw=none] at (1,0) (d) {w};
\node[draw=none] at (3,0) () {(bottom layer)};

\draw[->] (b) node[above] {} -- (a);
\draw[] (c) node[above] {} -- (a);
\draw[] (b) node[above] {} -- (d);
\end{tikzpicture}.$$

If the bottom strategies are not specified then the event includes all combinations of strategies on the bottom. For example, the event that the top strategy $x$ is replaced with strategy $y$, regardless of strategies on the bottom, is represented by,

$$\begin{tikzpicture}[baseline=3ex,scale=1]

\node[draw=none] at (0,1) (a) {x};
\node[draw=none] at (1,1) (b) {y};
\node[draw=none] at (3,1) () {(top layer)};

\node[draw=none] at (0,0) (c) {};
\node[draw=none] at (1,0) (d) {};
\node[draw=none] at (3,0) () {(bottom layer)};

\draw[->] (b) node[above] {} -- (a);
\draw[] (c) node[above] {} -- (a);
\draw[] (b) node[above] {} -- (d);
\end{tikzpicture}.$$

For the context interference spillover mode, if an individual with strategy $x$ is using strategy $v$ due to context interference, we denote her strategy as $x[ v ]$. Let the probability of an event $E$ be $\mathbb{P}(E)$. We are also using the facts that for the top layer, $x_{c}^{c} + x_{d}^{c} = q_c$, $x_{d}^{d} + x_{c}^{d} = q_d$, and that for the bottom layer, $x_{c}^{c} + x_{c}^{d} = p_c$, $x_{d}^{d} + x_{d}^{c} = p_d$. Finally, we let $\mathbb{E}(E)$ be the expected number of change in $C-C$ links when event $E$ occurs.

	% -----------------------------------------------------------------------------------------------------------------------------------%

\subsection{Neighbour Imitation Spillover}

At each discrete time step, a random individual is chosen. With probability $p$, the chosen individual does inter-layer neighbour imitation spillover updating. Otherwise, with probability $1-p$, she does intra-layer strategy updating. During neighbour imitation spillover updating, the top layer is chosen as the focal layer to {\bf receive} spillover strategies with probability $\alpha$. Otherwise, with probability $1-\alpha$ the bottom layer will be chosen as the focal layer to {\bf receive} spillover strategies from the top. \\ 

Then, the individual's strategy on the chosen focal layer might be replaced by the strategy of a random neighbour on the {\bf opposite} non-focal layer. Payoff comparison between this neighbour and the individual is then done, also on the opposite non-focal layer. This neighbour's strategy has a chance of being adopted by the individual on focal layer, given by the Fermi equation. Fig. 1B in the main text illustrates this process. We list the probabilities of all related events and configurations below.

		% ---- 		neighbour imitation probabilities for intra-layer strategy update 		-----

					% ----- 		Top Strategy Update 		-----

\[ \mathbb{P} \Bigg(
 \Bigg) = p \cdot \alpha \cdot p_c \cdot p_{c|c} \cdot \phi_{c \to c} \cdot \frac{x_{c}^{d}}{p_c}. \]

		% ---- 		neighbour imitation expected values 		-----

Let $n(x,y,z)$ be the number of $C$ strategies among the neighbours of the focal individual. Let $k_T$ be the average degree on the top layer, and $k_B$ be the average degree on the bottom layer. The list of expected change in number of $C-C$ links for relevant events and configurations are as follows.

	% ----- Top Strategy Update -----

\[ \mathbb{E} \Bigg(
%
 \Bigg) =  k_T \cdot q_{c|d}. \]

			% ------------- 	change in qc 		-----------------

Let $x,y,z$ be the strategies of the neighbours of the focal individual on the bottom layer. Let $u,v,w$ be strategies of the neighbours of the non-focal individual on the bottom layer. Then, the differential equations for the neighbour influence spillover mode are as follows.

\[ \dot{q}_c = \mathbb{P} \Bigg(
%
 \Bigg) \Bigg].	 \]

	% -----------------------------------------------------------------------------------------------------------------------------------%

\subsection{Self Comparison Spillover}

At each discrete time step, a random individual is chosen. With probability $p$, the chosen individual does inter-layer self comparison spillover updating. Otherwise, with probability $1-p$, she does intra-layer strategy updating. During self comparison spillover updating, the top layer is chosen as the focal layer to {\bf receive} spillover strategies with probability $\alpha$. Otherwise, with probability $1-\alpha$ the bottom layer will be chosen as the focal layer. \\ 

Then, the individual's strategy on the chosen focal layer might be replaced by the strategy of herself on the {\bf opposite} layer. The chance that this might happen depends on payoff comparison, calculated via the Fermi equation, with the exception that the payoff matrix for the iterated prisoners' dilemma game in the bottom layer is normalized by dividing each entry with the parameter $m$. We denote this normalized payoff matrix by,

$$ \begin{pmatrix} 
\hat{R}_B & \hat{S}_B \\
\hat{T}_B & \hat{P}_B
\end{pmatrix}.$$ 

Let $k_T$ be the average degree of individuals on the top layer. The average payoffs used by self comparison spillover on the top layer are

$$\hat{\pi}^{T}_c = k_T \cdot q_{c|c} \cdot R_T + k_T \cdot q_{d|c} \cdot S_T,$$

$$\hat{\pi}^{T}_d = k_T \cdot q_{c|d} \cdot T_T + k_T \cdot q_{d|d} \cdot P_T.$$

And the average normalized payoffs on the bottom layer, used for self comparison spillover updating, are

$$\hat{\pi}^{B}_c = n(w,x,y,z) \cdot \hat{R}_B + (k_B - n(w,x,y,z)) \cdot \hat{S}_B,$$

$$\hat{\pi}^{B}_d = n(w,x,y,z) \cdot \hat{T}_B + (k_B - n(w,x,y,z)) \cdot \hat{P}_B,$$

where $n(w,x,y,z)$ is the number of $C$ among neighbours strategies, of the focal individual, on the bottom layer. Then, the probabilities and expected changes in $C-C$ links of events that changes the system through self influence spillover update are as follows.

	% ----- 	Self Comparison Probabilities 	 -----

\[ \mathbb{P} \Bigg(
\begin{tikzpicture}[baseline=3ex,scale=1]

\node[draw=none] at (0,1) (a) {D};
\node[draw=none] at (0,0) (c) {C};

\draw[->] (c) node[above] {} -- (a);

\end{tikzpicture} \Bigg) = p \cdot \alpha \cdot F(\hat{\pi}^{B}_c,\hat{\pi}^{T}_d) \cdot x^{d}_{c}, 	\hspace{20pt}   \mathbb{P} \Bigg(
\begin{tikzpicture}[baseline=3ex,scale=1]

\node[draw=none] at (0,1) (a) {D};
\node[draw=none] at (0,0) (c) {C};

\draw[->] (a) node[above] {} -- (c);
\end{tikzpicture} \Bigg) = p \cdot (1-\alpha) \cdot F(\hat{\pi}^{T}_d,\hat{\pi}^{B}_c) \cdot x^{d}_{c}, \]

\[ \mathbb{P} \Bigg(
\begin{tikzpicture}[baseline=3ex,scale=1]

\node[draw=none] at (0,1) (a) {C};
\node[draw=none] at (0,0) (c) {D};

\draw[->] (c) node[above] {} -- (a);
\end{tikzpicture} \Bigg) =  p \cdot \alpha \cdot F(\hat{\pi}^{B}_d,\hat{\pi}^{T}_c) \cdot x^{c}_{d}, \hspace{20pt}   \mathbb{P} \Bigg(
\begin{tikzpicture}[baseline=3ex,scale=1]

\node[draw=none] at (0,1) (a) {C};
\node[draw=none] at (0,0) (c) {D};

\draw[->] (a) node[above] {} -- (c);
\end{tikzpicture} \Bigg) = p \cdot (1-\alpha) \cdot F(\hat{\pi}^{T}_c,\hat{\pi}^{B}_d) \cdot x^{c}_{d}, \]

	% ----- 	Self Comparison Expected Values 	 -----

\[ \mathbb{E} \Bigg(
\begin{tikzpicture}[baseline=3ex,scale=1]

\node[draw=none] at (0,1) (a) {C};
\node[draw=none] at (0,0) (c) {D};

\draw[->] (c) node[above] {} -- (a);

\end{tikzpicture} \Bigg) = -k_T \cdot q_{c|c}, 	\hspace{20pt}   \mathbb{E} \Bigg(
\begin{tikzpicture}[baseline=3ex,scale=1]

\node[draw=none] at (0,1) (a) {D};
\node[draw=none] at (0,0) (c) {C};

\draw[->] (c) node[above] {} -- (a);
\end{tikzpicture} \Bigg) = k_T \cdot q_{c|d}, \]

\[ \mathbb{E} \Bigg(
\begin{tikzpicture}[baseline=3ex,scale=1]

\node[draw=none] at (0,1) (a) {D};
\node[draw=none] at (0,0) (c) {C};

\draw[->] (a) node[above] {} -- (c);
\end{tikzpicture} \Bigg) = - k_B \cdot p_{c|c}, 	\hspace{20pt}   \mathbb{E} \Bigg(
\begin{tikzpicture}[baseline=3ex,scale=1]

\node[draw=none] at (0,1) (a) {C};
\node[draw=none] at (0,0) (c) {D};

\draw[->] (a) node[above] {} -- (c);
\end{tikzpicture} \Bigg) = k_B \cdot p_{c|d}, \]

where $F$ is the Fermi equation previously defined. Putting it all together, and using the information from the neighbour imitation section, we can write down the differential equations.

		% ----- q_c -----

% ----- first line = four intra-layer entries -----

\[ \dot{q}_c = \mathbb{P} \Bigg(
\begin{tikzpicture}[baseline=3ex,scale=1]

\node[draw=none] at (1,1) (b) {C};

\node[draw=none] at (0,1) (a) {D};
\node[draw=none] at (0,0) (c) {C};

\draw[->] (b) node[above] {} -- (a);
\draw[] (c) node[above] {} -- (a);
\end{tikzpicture} \Bigg)      +     \mathbb{P} \Bigg(
\begin{tikzpicture}[baseline=3ex,scale=1]

\node[draw=none] at (1,1) (b) {C};

\node[draw=none] at (0,1) (a) {D};
\node[draw=none] at (0,0) (c) {D};

\draw[->] (b) node[above] {} -- (a);
\draw[] (c) node[above] {} -- (a);
\end{tikzpicture} \Bigg)  -  \hspace{5pt}   \mathbb{P} \Bigg(
\begin{tikzpicture}[baseline=3ex,scale=1]

\node[draw=none] at (1,1) (b) {D};

\node[draw=none] at (0,1) (a) {C};
\node[draw=none] at (0,0) (c) {C};

\draw[->] (b) node[above] {} -- (a);
\draw[] (c) node[above] {} -- (a);
\end{tikzpicture} \Bigg)      -     \mathbb{P} \Bigg(
\begin{tikzpicture}[baseline=3ex,scale=1]

\node[draw=none] at (1,1) (b) {D};

\node[draw=none] at (0,1) (a) {C};
\node[draw=none] at (0,0) (c) {D};

\draw[->] (b) node[above] {} -- (a);
\draw[] (c) node[above] {} -- (a);
\end{tikzpicture} \Bigg)      \]

% ----- second line = two intra-layer entries -----

\[ + \sum_{x,y,z,w} \Bigg[ \mathbb{P} \Bigg(
\begin{tikzpicture}[baseline=3ex,scale=1]

\node[draw=none] at (0,1) (a) {D};
\node[draw=none] at (0,0) (c) {C};

\draw[->] (c) node[above] {} -- (a);
\end{tikzpicture} \Bigg)      -     \mathbb{P} \Bigg(
\begin{tikzpicture}[baseline=3ex,scale=1]

\node[draw=none] at (0,1) (a) {C};
\node[draw=none] at (0,0) (c) {D};

\draw[->] (c) node[above] {} -- (a);
\end{tikzpicture} \Bigg)  \Bigg]. \]

		% ----- q_cc -----

% ----- first line = four intra-layer entries -----

\[ \dot{q}_{cc} = \frac{2}{k_T} \cdot \Bigg\{ \mathbb{E} \Bigg(
\begin{tikzpicture}[baseline=3ex,scale=1]

\node[draw=none] at (1,1) (b) {D};

\node[draw=none] at (0,1) (a) {C};
\node[draw=none] at (0,0) (c) {C};

\draw[->] (b) node[above] {} -- (a);
\draw[] (c) node[above] {} -- (a);
\end{tikzpicture} \Bigg)     \cdot     \mathbb{P} \Bigg(
\begin{tikzpicture}[baseline=3ex,scale=1]

\node[draw=none] at (1,1) (b) {D};

\node[draw=none] at (0,1) (a) {C};
\node[draw=none] at (0,0) (c) {C};

\draw[->] (b) node[above] {} -- (a);
\draw[] (c) node[above] {} -- (a);
\end{tikzpicture} \Bigg)      +     \mathbb{E} \Bigg(
\begin{tikzpicture}[baseline=3ex,scale=1]

\node[draw=none] at (1,1) (b) {D};

\node[draw=none] at (0,1) (a) {C};
\node[draw=none] at (0,0) (c) {D};

\draw[->] (b) node[above] {} -- (a);
\draw[] (c) node[above] {} -- (a);
\end{tikzpicture} \Bigg)     \cdot     \mathbb{P} \Bigg(
\begin{tikzpicture}[baseline=3ex,scale=1]

\node[draw=none] at (1,1) (b) {D};

\node[draw=none] at (0,1) (a) {C};
\node[draw=none] at (0,0) (c) {D};

\draw[->] (b) node[above] {} -- (a);
\draw[] (c) node[above] {} -- (a);
\end{tikzpicture} \Bigg) \]   

\[ \hspace{30pt}    + \hspace{5pt}  \mathbb{E} \Bigg(
\begin{tikzpicture}[baseline=3ex,scale=1]

\node[draw=none] at (1,1) (b) {C};

\node[draw=none] at (0,1) (a) {D};
\node[draw=none] at (0,0) (c) {C};

\draw[->] (b) node[above] {} -- (a);
\draw[] (c) node[above] {} -- (a);
\end{tikzpicture} \Bigg)     \cdot     \mathbb{P} \Bigg(
\begin{tikzpicture}[baseline=3ex,scale=1]

\node[draw=none] at (1,1) (b) {C};

\node[draw=none] at (0,1) (a) {D};
\node[draw=none] at (0,0) (c) {C};

\draw[->] (b) node[above] {} -- (a);
\draw[] (c) node[above] {} -- (a);
\end{tikzpicture} \Bigg)       +      \mathbb{E} \Bigg(
\begin{tikzpicture}[baseline=3ex,scale=1]

\node[draw=none] at (1,1) (b) {C};

\node[draw=none] at (0,1) (a) {D};
\node[draw=none] at (0,0) (c) {D};

\draw[->] (b) node[above] {} -- (a);
\draw[] (c) node[above] {} -- (a);
\end{tikzpicture} \Bigg)     \cdot     \mathbb{P} \Bigg(
\begin{tikzpicture}[baseline=3ex,scale=1]

\node[draw=none] at (1,1) (b) {C};

\node[draw=none] at (0,1) (a) {D};
\node[draw=none] at (0,0) (c) {D};

\draw[->] (b) node[above] {} -- (a);
\draw[] (c) node[above] {} -- (a);
\end{tikzpicture} \Bigg)    \]

% ----- second line = two inter-layer entries -----

\[	+ \sum_{x,y,z,w} \Bigg[ \mathbb{E} \Bigg(
\begin{tikzpicture}[baseline=3ex,scale=1]

\node[draw=none] at (0,1) (a) {C};
\node[draw=none] at (0,0) (c) {D};

\draw[->] (c) node[above] {} -- (a);

\end{tikzpicture} \Bigg)     \cdot      \mathbb{P} \Bigg(
\begin{tikzpicture}[baseline=3ex,scale=1]

\node[draw=none] at (0,1) (a) {C};
\node[draw=none] at (0,0) (c) {D};

\draw[->] (c) node[above] {} -- (a);

\end{tikzpicture} \Bigg) 	+	\mathbb{E} \Bigg(
\begin{tikzpicture}[baseline=3ex,scale=1]

\node[draw=none] at (0,1) (a) {D};
\node[draw=none] at (0,0) (c) {C};

\draw[->] (c) node[above] {} -- (a);

\end{tikzpicture} \Bigg)      \cdot     \mathbb{P} \Bigg(
\begin{tikzpicture}[baseline=3ex,scale=1]

\node[draw=none] at (0,1) (a) {D};
\node[draw=none] at (0,0) (c) {C};

\draw[->] (c) node[above] {} -- (a);
\end{tikzpicture} \Bigg)	\Bigg]	\Bigg\}. \]

	% ------ Bottom Layer ------

		% ----- p_c -----
		
% ----- first line = four intra-layer entries -----

\[ \dot{p}_c = \sum_{x,y,z,u,v,w}  \Bigg[  \mathbb{P} \Bigg(
\begin{tikzpicture}[baseline=3ex,scale=1]

\node[draw=none] at (0,1) (a) {C};
\node[draw=none] at (0,0) (c) {D};

\node[draw=none] at (1,0) (b) {C};

\draw[->] (b) node[above] {} -- (c);
\draw[] (c) node[above] {} -- (a);

\end{tikzpicture} \Bigg)      +     \mathbb{P} \Bigg(
\begin{tikzpicture}[baseline=3ex,scale=1]

\node[draw=none] at (0,1) (a) {D};
\node[draw=none] at (0,0) (c) {D};

\node[draw=none] at (1,0) (b) {C};

\draw[->] (b) node[above] {} -- (c);
\draw[] (c) node[above] {} -- (a);
\end{tikzpicture} \Bigg)	 -  \hspace{5pt}   \mathbb{P} \Bigg(
\begin{tikzpicture}[baseline=3ex,scale=1]

\node[draw=none] at (0,1) (a) {C};
\node[draw=none] at (0,0) (c) {C};

\node[draw=none] at (1,0) (b) {D};

\draw[->] (b) node[above] {} -- (c);
\draw[] (c) node[above] {} -- (a);
\end{tikzpicture} \Bigg)      -     \mathbb{P} \Bigg(
\begin{tikzpicture}[baseline=3ex,scale=1]

\node[draw=none] at (0,1) (a) {D};
\node[draw=none] at (0,0) (c) {C};

\node[draw=none] at (1,0) (b) {D};

\draw[->] (b) node[above] {} -- (c);
\draw[] (c) node[above] {} -- (a);
\end{tikzpicture} \Bigg)  \Bigg]		\]

% ----- second line = two inter-layer entries -----

\[	+ \sum_{x,y,z,w} \Bigg[ \mathbb{P} \Bigg(
\begin{tikzpicture}[baseline=3ex,scale=1]

\node[draw=none] at (0,1) (a) {C};
\node[draw=none] at (0,0) (c) {D};

\draw[->] (a) node[above] {} -- (c);
\end{tikzpicture} \Bigg)       -      \mathbb{P} \Bigg(
\begin{tikzpicture}[baseline=3ex,scale=1]

\node[draw=none] at (0,1) (a) {D};
\node[draw=none] at (0,0) (c) {C};

\draw[->] (a) node[above] {} -- (c);
\end{tikzpicture} \Bigg) \Bigg]. \]

% ----- first and second lines = four intra-layer entries -----

\[ \dot{p}_{cc} =\frac{2}{k_T} \cdot \Bigg\{ \sum_{x,y,z,u,v,w}  \Bigg[ \mathbb{E} \Bigg(
\begin{tikzpicture}[baseline=3ex,scale=1]

\node[draw=none] at (1,0) (b) {D};

\node[draw=none] at (0,1) (a) {D};
\node[draw=none] at (0,0) (c) {C};

\draw[] (c) node[above] {} -- (a);
\draw[->] (b) node[above] {} -- (c);
\end{tikzpicture} \Bigg)     \cdot     \mathbb{P} \Bigg(
\begin{tikzpicture}[baseline=3ex,scale=1]

\node[draw=none] at (1,0) (b) {D};

\node[draw=none] at (0,1) (a) {D};
\node[draw=none] at (0,0) (c) {C};

\draw[] (c) node[above] {} -- (a);
\draw[->] (b) node[above] {} -- (c);
\end{tikzpicture} \Bigg)      +     \mathbb{E} \Bigg(
\begin{tikzpicture}[baseline=3ex,scale=1]

\node[draw=none] at (1,0) (b) {D};

\node[draw=none] at (0,1) (a) {C};
\node[draw=none] at (0,0) (c) {C};

\draw[] (c) node[above] {} -- (a);
\draw[->] (b) node[above] {} -- (c);

\end{tikzpicture} \Bigg)     \cdot     \mathbb{P} \Bigg(
\begin{tikzpicture}[baseline=3ex,scale=1]

\node[draw=none] at (1,0) (b) {D};

\node[draw=none] at (0,1) (a) {C};
\node[draw=none] at (0,0) (c) {C};

\draw[] (c) node[above] {} -- (a);
\draw[->] (b) node[above] {} -- (c);

\end{tikzpicture} \Bigg) \]   

\[ \hspace{30pt}    + \hspace{5pt}  \mathbb{E} \Bigg(
\begin{tikzpicture}[baseline=3ex,scale=1]

\node[draw=none] at (1,0) (b) {C};

\node[draw=none] at (0,1) (a) {C};
\node[draw=none] at (0,0) (c) {D};

\draw[] (c) node[above] {} -- (a);
\draw[->] (b) node[above] {} -- (c);

\end{tikzpicture} \Bigg)     \cdot     \mathbb{P} \Bigg(
\begin{tikzpicture}[baseline=3ex,scale=1]

\node[draw=none] at (1,0) (b) {C};

\node[draw=none] at (0,1) (a) {C};
\node[draw=none] at (0,0) (c) {D};

\draw[] (c) node[above] {} -- (a);
\draw[->] (b) node[above] {} -- (c);

\end{tikzpicture} \Bigg)       +      \mathbb{E} \Bigg(
\begin{tikzpicture}[baseline=3ex,scale=1]

\node[draw=none] at (1,0) (b) {C};

\node[draw=none] at (0,1) (a) {D};
\node[draw=none] at (0,0) (c) {D};

\draw[] (c) node[above] {} -- (a);
\draw[->] (b) node[above] {} -- (c);

\end{tikzpicture} \Bigg)     \cdot     \mathbb{P} \Bigg(
\begin{tikzpicture}[baseline=3ex,scale=1]

\node[draw=none] at (1,0) (b) {C};

\node[draw=none] at (0,1) (a) {D};
\node[draw=none] at (0,0) (c) {D};

\draw[] (c) node[above] {} -- (a);
\draw[->] (b) node[above] {} -- (c);

\end{tikzpicture} \Bigg)	\Bigg]	 \]

% ----- second line = two inter-layer entries -----

\[ \hspace{30pt}    + \sum_{x,y,z,w} \Bigg[     \mathbb{E} \Bigg(
\begin{tikzpicture}[baseline=3ex,scale=1]

\node[draw=none] at (0,1) (a) {C};
\node[draw=none] at (0,0) (c) {D};

\draw[->] (a) node[above] {} -- (c);

\end{tikzpicture} \Bigg)      \cdot     \mathbb{P} \Bigg(
\begin{tikzpicture}[baseline=3ex,scale=1]

\node[draw=none] at (0,1) (a) {C};
\node[draw=none] at (0,0) (c) {D};

\draw[->] (a) node[above] {} -- (c);

\end{tikzpicture} \Bigg)     +      \mathbb{E} \Bigg(
\begin{tikzpicture}[baseline=3ex,scale=1]

\node[draw=none] at (0,1) (a) {D};
\node[draw=none] at (0,0) (c) {C};

\draw[->] (a) node[above] {} -- (c);

\end{tikzpicture} \Bigg)       \cdot     \mathbb{P} \Bigg(
\begin{tikzpicture}[baseline=3ex,scale=1]

\node[draw=none] at (0,1) (a) {D};
\node[draw=none] at (0,0) (c) {C};

\draw[->] (a) node[above] {} -- (c);

\end{tikzpicture} \Bigg)	\Bigg]	\Bigg\}.	 \]

\vspace{20pt}	% ----- End of p_cc -----

	% ------ Inter-layer Proportions aXb ------

		% ----- cXc -----

% ----- first and second lines: four intra-layer entries -----

\[ \dot{x}^{c}_{c} = \hspace{5pt}  \sum_{x,y,z,u,v,w} \Bigg[ 
\mathbb{P} \Bigg(
\begin{tikzpicture}[baseline=3ex,scale=1]

\node[draw=none] at (1,0) (b) {C};

\node[draw=none] at (0,1) (a) {C};
\node[draw=none] at (0,0) (c) {D};

\draw[] (c) node[above] {} -- (a);
\draw[->] (b) node[above] {} -- (c);

\end{tikzpicture} \Bigg)	-	\mathbb{P} \Bigg(
\begin{tikzpicture}[baseline=3ex,scale=1]

\node[draw=none] at (1,0) (b) {D};

\node[draw=none] at (0,1) (a) {C};
\node[draw=none] at (0,0) (c) {C};

\draw[] (c) node[above] {} -- (a);
\draw[->] (b) node[above] {} -- (c);

\end{tikzpicture} \Bigg)  	\Bigg]    \]

	% ----- Line Break -----

\[ 	\hspace{30pt}    +  \hspace{5pt}      	\mathbb{P} \Bigg(
\begin{tikzpicture}[baseline=3ex,scale=1]

\node[draw=none] at (1,1) (b) {C};

\node[draw=none] at (0,1) (a) {D};
\node[draw=none] at (0,0) (c) {C};

\draw[->] (b) node[above] {} -- (a);
\draw[] (c) node[above] {} -- (a);

\end{tikzpicture} \Bigg) 	-      \mathbb{P} \Bigg(
\begin{tikzpicture}[baseline=3ex,scale=1]

\node[draw=none] at (1,1) (b) {D};

\node[draw=none] at (0,1) (a) {C};
\node[draw=none] at (0,0) (c) {C};

\draw[->] (b) node[above] {} -- (a);
\draw[] (c) node[above] {} -- (a);

\end{tikzpicture} \Bigg)	 \]    

% ----- third line: two inter-layer entries -----

\[ + \sum_{x,y,z,w} \Bigg[  \mathbb{P} \Bigg(
\begin{tikzpicture}[baseline=3ex,scale=1]

\node[draw=none] at (0,1) (a) {D};
\node[draw=none] at (0,0) (c) {C};

\draw[->] (c) node[above] {} -- (a);

\end{tikzpicture} \Bigg)      +     \mathbb{P} \Bigg(
\begin{tikzpicture}[baseline=3ex,scale=1]

\node[draw=none] at (0,1) (a) {C};
\node[draw=none] at (0,0) (c) {D};

\draw[->] (a) node[above] {} -- (c);

\end{tikzpicture} \Bigg)	\Bigg].   \]

% ----- End of cXc -----

		% ----- x^d_c -----

% ----- first and second lines: four intra-layer entries -----

\[ \dot{x}^{d}_{c} = \hspace{5pt} \sum_{x,y,z,u,v,w} \Bigg[     \mathbb{P} \Bigg(
\begin{tikzpicture}[baseline=3ex,scale=1]

\node[draw=none] at (1,0) (b) {C};

\node[draw=none] at (0,1) (a) {D};
\node[draw=none] at (0,0) (c) {D};

\draw[] (c) node[above] {} -- (a);
\draw[->] (b) node[above] {} -- (c);

\end{tikzpicture} \Bigg) 	- 	\mathbb{P} \Bigg(
\begin{tikzpicture}[baseline=3ex,scale=1]

\node[draw=none] at (1,0) (b) {D};

\node[draw=none] at (0,1) (a) {D};
\node[draw=none] at (0,0) (c) {C};

\draw[] (c) node[above] {} -- (a);
\draw[->] (b) node[above] {} -- (c);

\end{tikzpicture} \Bigg) 	\Bigg] 	\]

	% ----- Line Break -----

\[     \hspace{30pt}    +  \hspace{5pt}  \mathbb{P} \Bigg(
\begin{tikzpicture}[baseline=3ex,scale=1]

\node[draw=none] at (1,1) (b) {D};

\node[draw=none] at (0,1) (a) {C};
\node[draw=none] at (0,0) (c) {C};

\draw[->] (b) node[above] {} -- (a);
\draw[] (c) node[above] {} -- (a);

\end{tikzpicture} \Bigg)  -  \hspace{5pt}		\mathbb{P} \Bigg(
\begin{tikzpicture}[baseline=3ex,scale=1]

\node[draw=none] at (1,1) (b) {C};

\node[draw=none] at (0,1) (a) {D};
\node[draw=none] at (0,0) (c) {C};

\draw[->] (b) node[above] {} -- (a);
\draw[] (c) node[above] {} -- (a);

\end{tikzpicture} \Bigg)	\]

% ----- third line: two inter-layer entries -----

\[ 	- \sum_{x,y,z,w} \Bigg[ \mathbb{P} \Bigg(
\begin{tikzpicture}[baseline=3ex,scale=1]

\node[draw=none] at (0,1) (a) {D};
\node[draw=none] at (0,0) (c) {C};

\draw[->] (c) node[above] {} -- (a);

\end{tikzpicture} \Bigg)      +     \mathbb{P} \Bigg(
\begin{tikzpicture}[baseline=3ex,scale=1]

\node[draw=none] at (0,1) (a) {D};
\node[draw=none] at (0,0) (c) {C};

\draw[->] (a) node[above] {} -- (c);

\end{tikzpicture} \Bigg)	\Bigg].    \]

	% -----------------------------------------------------------------------------------------------------------------------------------%

\subsection{Context Interference Spillover} 				% ----- 	context interference	-----

At each discrete time step, a random layer and a random individual on that layer is chosen as the focal individual for strategy updating.  A random neighbour of this individual is then chosen and their payoffs are compared via the Fermi equation. When calculating payoffs, with probability $p$, the focal individual experiences context interference and uses her strategy from the bottom layer with probability $\alpha$ and her strategy from the top layer with probability $1-\alpha$. Independently, also with probability $p$, the chosen neighbour can experience the same context interference. Strategy updating is then carried out with these strategies. The differential equation for the change in fraction of $C$ in the top layer is,

			% ----- 	change in q_c 	-----

\[ \dot{q}_c = \mathbb{P} \Bigg(
\begin{tikzpicture}[baseline=3ex,scale=1]

\node[draw=none] at (0,1) (a) {D}; % 1st colum
\node[draw=none] at (0,0) (b) {};

\node[draw=none] at (1,1) (c) {C}; % 2nd column
\node[draw=none] at (1,0) (d) {};

\draw[->] (c) node[above] {} -- (a);
\draw[] (a) node[above] {} -- (b);
\draw[] (c) node[above] {} -- (d);

\end{tikzpicture} \Bigg)	+ 	\mathbb{P} \Bigg(
\begin{tikzpicture}[baseline=3ex,scale=1]

\node[draw=none] at (0,1) (a) {D}; % 1st colum
\node[draw=none] at (0,0) (b) {};

\node[draw=none] at (1,1) (c) {D[C]}; % 2nd column
\node[draw=none] at (1,0) (d) {C};

\draw[->] (c) node[above] {} -- (a);
\draw[] (a) node[above] {} -- (b);
\draw[] (c) node[above] {} -- (d);

\end{tikzpicture} \Bigg)	+ 	\mathbb{P} \Bigg(
\begin{tikzpicture}[baseline=3ex,scale=1]

\node[draw=none] at (0,1) (a) {D[C]}; % 1st colum
\node[draw=none] at (0,0) (b) {C};

\node[draw=none] at (1,1) (c) {C}; % 2nd column
\node[draw=none] at (1,0) (d) {};

\draw[->] (c) node[above] {} -- (a);
\draw[] (a) node[above] {} -- (b);
\draw[] (c) node[above] {} -- (d);

\end{tikzpicture} \Bigg)	+ 	\mathbb{P} \Bigg(
\begin{tikzpicture}[baseline=3ex,scale=1]

\node[draw=none] at (0,1) (a) {D[C]}; % 1st colum
\node[draw=none] at (0,0) (b) {C};

\node[draw=none] at (1.3,1) (c) {D[C]}; % 2nd column
\node[draw=none] at (1.3,0) (d) {C};

\draw[->] (c) node[above] {} -- (a);
\draw[] (a) node[above] {} -- (b);
\draw[] (c) node[above] {} -- (d);

\end{tikzpicture} \Bigg)      \]

												% ----- 		2nd line of qc	-----

\[	\hspace{15pt}	- 	 \mathbb{P} \Bigg(
\begin{tikzpicture}[baseline=3ex,scale=1]

\node[draw=none] at (0,1) (a) {C}; % 1st colum
\node[draw=none] at (0,0) (b) {};

\node[draw=none] at (1,1) (c) {D}; % 2nd column
\node[draw=none] at (1,0) (d) {};

\draw[->] (c) node[above] {} -- (a);
\draw[] (a) node[above] {} -- (b);
\draw[] (c) node[above] {} -- (d);

\end{tikzpicture} \Bigg)	- 	\mathbb{P} \Bigg(
\begin{tikzpicture}[baseline=3ex,scale=1]

\node[draw=none] at (0,1) (a) {C}; % 1st colum
\node[draw=none] at (0,0) (b) {};

\node[draw=none] at (1,1) (c) {C[D]}; % 2nd column
\node[draw=none] at (1,0) (d) {D};

\draw[->] (c) node[above] {} -- (a);
\draw[] (a) node[above] {} -- (b);
\draw[] (c) node[above] {} -- (d);

\end{tikzpicture} \Bigg)	- 	\mathbb{P} \Bigg(
\begin{tikzpicture}[baseline=3ex,scale=1]

\node[draw=none] at (0,1) (a) {C[D]}; % 1st colum
\node[draw=none] at (0,0) (b) {D};

\node[draw=none] at (1,1) (c) {D}; % 2nd column
\node[draw=none] at (1,0) (d) {};

\draw[->] (c) node[above] {} -- (a);
\draw[] (a) node[above] {} -- (b);
\draw[] (c) node[above] {} -- (d);

\end{tikzpicture} \Bigg)	- 	\mathbb{P} \Bigg(
\begin{tikzpicture}[baseline=3ex,scale=1]

\node[draw=none] at (0,1) (a) {C[D]}; % 1st colum
\node[draw=none] at (0,0) (b) {D};

\node[draw=none] at (1.3,1) (c) {C[D]}; % 2nd column
\node[draw=none] at (1.3,0) (d) {D};

\draw[->] (c) node[above] {} -- (a);
\draw[] (a) node[above] {} -- (b);
\draw[] (c) node[above] {} -- (d);

\end{tikzpicture} \Bigg).      \]

Detailed equations for each of the terms in the sum are as follows.

			% ---- details for terms in dot q_c ----

\[ \mathbb{P} \Bigg(						% ----- 1st term -----
\begin{tikzpicture}[baseline=3ex,scale=1]

\node[draw=none] at (0,1) (a) {D}; % 1st colum
\node[draw=none] at (0,0) (b) {};

\node[draw=none] at (1,1) (c) {C}; % 2nd column
\node[draw=none] at (1,0) (d) {};

\draw[->] (c) node[above] {} -- (a);
\draw[] (a) node[above] {} -- (b);
\draw[] (c) node[above] {} -- (d);

\end{tikzpicture} \Bigg)		=	 	 \mathbb{P} \Bigg(
\begin{tikzpicture}[baseline=3ex,scale=1]

\node[draw=none] at (0,1) (a) {D}; % 1st colum
\node[draw=none] at (0,0) (b) {C};

\node[draw=none] at (1,1) (c) {C}; % 2nd column
\node[draw=none] at (1,0) (d) {C};

\draw[->] (c) node[above] {} -- (a);
\draw[] (a) node[above] {} -- (b);
\draw[] (c) node[above] {} -- (d);

\end{tikzpicture} \Bigg)	+ 	\mathbb{P} \Bigg(
\begin{tikzpicture}[baseline=3ex,scale=1]

\node[draw=none] at (0,1) (a) {D}; % 1st colum
\node[draw=none] at (0,0) (b) {C};

\node[draw=none] at (1,1) (c) {C}; % 2nd column
\node[draw=none] at (1,0) (d) {D};

\draw[->] (c) node[above] {} -- (a);
\draw[] (a) node[above] {} -- (b);
\draw[] (c) node[above] {} -- (d);

\end{tikzpicture} \Bigg)	+ 	\mathbb{P} \Bigg(
\begin{tikzpicture}[baseline=3ex,scale=1]

\node[draw=none] at (0,1) (a) {D}; % 1st colum
\node[draw=none] at (0,0) (b) {D};

\node[draw=none] at (1,1) (c) {C}; % 2nd column
\node[draw=none] at (1,0) (d) {C};

\draw[->] (c) node[above] {} -- (a);
\draw[] (a) node[above] {} -- (b);
\draw[] (c) node[above] {} -- (d);

\end{tikzpicture} \Bigg)	+ 	\mathbb{P} \Bigg(
\begin{tikzpicture}[baseline=3ex,scale=1]

\node[draw=none] at (0,1) (a) {D}; % 1st colum
\node[draw=none] at (0,0) (b) {D};

\node[draw=none] at (1,1) (c) {C}; % 2nd column
\node[draw=none] at (1,0) (d) {D};

\draw[->] (c) node[above] {} -- (a);
\draw[] (a) node[above] {} -- (b);
\draw[] (c) node[above] {} -- (d);

\end{tikzpicture} \Bigg)\]

\[ = \frac{1}{2} \cdot W_{c \to d} \cdot q_d \cdot q_{c|d} \cdot \Bigg(  
\frac{X^d_c}{q_d} \cdot \frac{X^c_c}{q_c} \cdot (1-p + (1-\alpha) \cdot p) 
+ \frac{X^d_c}{q_d} \cdot \frac{X^c_d}{q_c} \cdot (1-p + (1-\alpha) \cdot p) \cdot (1-p + (1-\alpha) \cdot p) \]

\[
+ \frac{X^d_d}{q_d} \cdot \frac{X^c_c}{q_c}
+ \frac{X^d_d}{q_d} \cdot \frac{X^c_d}{q_c} \cdot (1-p + (1-\alpha) \cdot p) 
\Bigg). \]

\vspace{20pt}

\[ \mathbb{P} \Bigg(						% ----- 2nd term -----
\begin{tikzpicture}[baseline=3ex,scale=1]

\node[draw=none] at (0,1) (a) {D}; % 1st colum
\node[draw=none] at (0,0) (b) {};

\node[draw=none] at (1,1) (c) {D[C]}; % 2nd column
\node[draw=none] at (1,0) (d) {C};

\draw[->] (c) node[above] {} -- (a);
\draw[] (a) node[above] {} -- (b);
\draw[] (c) node[above] {} -- (d);

\end{tikzpicture} \Bigg)		=	 	 \mathbb{P} \Bigg(
\begin{tikzpicture}[baseline=3ex,scale=1]

\node[draw=none] at (0,1) (a) {D}; % 1st colum
\node[draw=none] at (0,0) (b) {C};

\node[draw=none] at (1,1) (c) {D[C]}; % 2nd column
\node[draw=none] at (1,0) (d) {C};

\draw[->] (c) node[above] {} -- (a);
\draw[] (a) node[above] {} -- (b);
\draw[] (c) node[above] {} -- (d);

\end{tikzpicture} \Bigg)	+ 	\mathbb{P} \Bigg(
\begin{tikzpicture}[baseline=3ex,scale=1]

\node[draw=none] at (0,1) (a) {D}; % 1st colum
\node[draw=none] at (0,0) (b) {D};

\node[draw=none] at (1,1) (c) {D[C]}; % 2nd column
\node[draw=none] at (1,0) (d) {C};

\draw[->] (c) node[above] {} -- (a);
\draw[] (a) node[above] {} -- (b);
\draw[] (c) node[above] {} -- (d);

\end{tikzpicture} \Bigg)\]

\[ = \frac{1}{2} \cdot W_{c \to d} \cdot q_d \cdot q_{d|d} \cdot \Bigg(  
\frac{X^d_c}{q_d} \cdot \frac{X^d_c}{q_d} \cdot (1-p + (1-\alpha) \cdot p) \cdot (\alpha \cdot p) 
+ \frac{X^d_d}{q_d} \cdot \frac{X^d_c}{q_c} \cdot (\alpha \cdot p) 
\Bigg). \]

\vspace{20pt}

\[ \mathbb{P} \Bigg(						% ----- 3rd term -----
\begin{tikzpicture}[baseline=3ex,scale=1]

\node[draw=none] at (0,1) (a) {D[C]}; % 1st colum
\node[draw=none] at (0,0) (b) {C};

\node[draw=none] at (1,1) (c) {C}; % 2nd column
\node[draw=none] at (1,0) (d) {};

\draw[->] (c) node[above] {} -- (a);
\draw[] (a) node[above] {} -- (b);
\draw[] (c) node[above] {} -- (d);

\end{tikzpicture} \Bigg)		=	 	 \mathbb{P} \Bigg(
\begin{tikzpicture}[baseline=3ex,scale=1]

\node[draw=none] at (0,1) (a) {D[C]}; % 1st colum
\node[draw=none] at (0,0) (b) {C};

\node[draw=none] at (1,1) (c) {C}; % 2nd column
\node[draw=none] at (1,0) (d) {C};

\draw[->] (c) node[above] {} -- (a);
\draw[] (a) node[above] {} -- (b);
\draw[] (c) node[above] {} -- (d);

\end{tikzpicture} \Bigg)	+ 	\mathbb{P} \Bigg(
\begin{tikzpicture}[baseline=3ex,scale=1]

\node[draw=none] at (0,1) (a) {D[C]}; % 1st colum
\node[draw=none] at (0,0) (b) {C};

\node[draw=none] at (1,1) (c) {C}; % 2nd column
\node[draw=none] at (1,0) (d) {D};

\draw[->] (c) node[above] {} -- (a);
\draw[] (a) node[above] {} -- (b);
\draw[] (c) node[above] {} -- (d);

\end{tikzpicture} \Bigg)\]

\[ = \frac{1}{2} \cdot W_{c \to c} \cdot q_d \cdot q_{c|d} \cdot \frac{X^d_c}{q_d} \cdot (\alpha \cdot p)  \cdot \Bigg(  
\frac{X^c_c}{q_c}
+ \frac{X^c_d}{q_c} \cdot (1-p + (1-\alpha) \cdot p) 
\Bigg). \]

\vspace{20pt}

\[ \mathbb{P} \Bigg(						% ----- 4th term -----
\begin{tikzpicture}[baseline=3ex,scale=1]

\node[draw=none] at (0,1) (a) {D[C]}; % 1st colum
\node[draw=none] at (0,0) (b) {C};

\node[draw=none] at (1.3,1) (c) {D[C]}; % 2nd column
\node[draw=none] at (1.3,0) (d) {C};

\draw[->] (c) node[above] {} -- (a);
\draw[] (a) node[above] {} -- (b);
\draw[] (c) node[above] {} -- (d);

\end{tikzpicture} \Bigg)
= \frac{1}{2} \cdot W_{c \to c} \cdot q_d \cdot q_{d|d} \cdot \Bigg(  
\frac{X^d_c}{q_d} \cdot \frac{X^d_c}{q_d} \cdot (\alpha \cdot p)  \cdot (\alpha \cdot p)  
\Bigg). \]

\vspace{20pt}

\[ \mathbb{P} \Bigg(						% ----- 5th term -----
\begin{tikzpicture}[baseline=3ex,scale=1]

\node[draw=none] at (0,1) (a) {C}; % 1st colum
\node[draw=none] at (0,0) (b) {};

\node[draw=none] at (1,1) (c) {D}; % 2nd column
\node[draw=none] at (1,0) (d) {};

\draw[->] (c) node[above] {} -- (a);
\draw[] (a) node[above] {} -- (b);
\draw[] (c) node[above] {} -- (d);

\end{tikzpicture} \Bigg)		=	 	 \mathbb{P} \Bigg(
\begin{tikzpicture}[baseline=3ex,scale=1]

\node[draw=none] at (0,1) (a) {C}; % 1st colum
\node[draw=none] at (0,0) (b) {C};

\node[draw=none] at (1,1) (c) {D}; % 2nd column
\node[draw=none] at (1,0) (d) {C};

\draw[->] (c) node[above] {} -- (a);
\draw[] (a) node[above] {} -- (b);
\draw[] (c) node[above] {} -- (d);

\end{tikzpicture} \Bigg)	+ 	\mathbb{P} \Bigg(
\begin{tikzpicture}[baseline=3ex,scale=1]

\node[draw=none] at (0,1) (a) {C}; % 1st colum
\node[draw=none] at (0,0) (b) {C};

\node[draw=none] at (1,1) (c) {D}; % 2nd column
\node[draw=none] at (1,0) (d) {D};

\draw[->] (c) node[above] {} -- (a);
\draw[] (a) node[above] {} -- (b);
\draw[] (c) node[above] {} -- (d);

\end{tikzpicture} \Bigg)	+ 	\mathbb{P} \Bigg(
\begin{tikzpicture}[baseline=3ex,scale=1]

\node[draw=none] at (0,1) (a) {C}; % 1st colum
\node[draw=none] at (0,0) (b) {D};

\node[draw=none] at (1,1) (c) {D}; % 2nd column
\node[draw=none] at (1,0) (d) {C};

\draw[->] (c) node[above] {} -- (a);
\draw[] (a) node[above] {} -- (b);
\draw[] (c) node[above] {} -- (d);

\end{tikzpicture} \Bigg)	+ 	\mathbb{P} \Bigg(
\begin{tikzpicture}[baseline=3ex,scale=1]

\node[draw=none] at (0,1) (a) {C}; % 1st colum
\node[draw=none] at (0,0) (b) {D};

\node[draw=none] at (1,1) (c) {D}; % 2nd column
\node[draw=none] at (1,0) (d) {D};

\draw[->] (c) node[above] {} -- (a);
\draw[] (a) node[above] {} -- (b);
\draw[] (c) node[above] {} -- (d);

\end{tikzpicture} \Bigg)\]

\[ = \frac{1}{2} \cdot W_{d \to c} \cdot q_c \cdot q_{d|c} \cdot \Bigg(  
\frac{X^c_c}{q_c} \cdot \frac{X^d_c}{q_d} \cdot (1-p + (1-\alpha) \cdot p) 
+ \frac{X^c_c}{q_c} \cdot \frac{X^d_d}{q_d} \]

\[
+ \frac{X^c_d}{q_c} \cdot \frac{X^d_c}{q_d} \cdot (1-p + (1-\alpha) \cdot p) \cdot (1-p + (1-\alpha) \cdot p) 
+ \frac{X^c_d}{q_c} \cdot \frac{X^d_d}{q_d} \cdot (1-p + (1-\alpha) \cdot p) 
\Bigg). \]

\[ \mathbb{P} \Bigg(						% ----- 6th term -----
\begin{tikzpicture}[baseline=3ex,scale=1]

\node[draw=none] at (0,1) (a) {C}; % 1st colum
\node[draw=none] at (0,0) (b) {};

\node[draw=none] at (1,1) (c) {C[D]}; % 2nd column
\node[draw=none] at (1,0) (d) {D};

\draw[->] (c) node[above] {} -- (a);
\draw[] (a) node[above] {} -- (b);
\draw[] (c) node[above] {} -- (d);

\end{tikzpicture} \Bigg)		=	 	 \mathbb{P} \Bigg(
\begin{tikzpicture}[baseline=3ex,scale=1]

\node[draw=none] at (0,1) (a) {C}; % 1st colum
\node[draw=none] at (0,0) (b) {C};

\node[draw=none] at (1,1) (c) {C[D]}; % 2nd column
\node[draw=none] at (1,0) (d) {D};

\draw[->] (c) node[above] {} -- (a);
\draw[] (a) node[above] {} -- (b);
\draw[] (c) node[above] {} -- (d);

\end{tikzpicture} \Bigg)	+ 	\mathbb{P} \Bigg(
\begin{tikzpicture}[baseline=3ex,scale=1]

\node[draw=none] at (0,1) (a) {C}; % 1st colum
\node[draw=none] at (0,0) (b) {D};

\node[draw=none] at (1,1) (c) {C[D]}; % 2nd column
\node[draw=none] at (1,0) (d) {D};

\draw[->] (c) node[above] {} -- (a);
\draw[] (a) node[above] {} -- (b);
\draw[] (c) node[above] {} -- (d);

\end{tikzpicture} \Bigg)\]

\[ = \frac{1}{2} \cdot W_{d \to c} \cdot q_c \cdot q_{c|c}  \cdot \frac{X^c_d}{q_c} \cdot (\alpha \cdot p)  \cdot \Bigg(  
\frac{X^c_c}{q_c}
+ \frac{X^c_d}{q_c} \cdot (1-p + (1-\alpha) \cdot p)
\Bigg). \]

\vspace{20pt}

\[ \mathbb{P} \Bigg(						% ----- 7th term -----
\begin{tikzpicture}[baseline=3ex,scale=1]

\node[draw=none] at (0,1) (a) {C[D]}; % 1st colum
\node[draw=none] at (0,0) (b) {D};

\node[draw=none] at (1,1) (c) {D}; % 2nd column
\node[draw=none] at (1,0) (d) {};

\draw[->] (c) node[above] {} -- (a);
\draw[] (a) node[above] {} -- (b);
\draw[] (c) node[above] {} -- (d);

\end{tikzpicture} \Bigg)		=	 	 \mathbb{P} \Bigg(
\begin{tikzpicture}[baseline=3ex,scale=1]

\node[draw=none] at (0,1) (a) {C[D]}; % 1st colum
\node[draw=none] at (0,0) (b) {D};

\node[draw=none] at (1,1) (c) {D}; % 2nd column
\node[draw=none] at (1,0) (d) {C};

\draw[->] (c) node[above] {} -- (a);
\draw[] (a) node[above] {} -- (b);
\draw[] (c) node[above] {} -- (d);

\end{tikzpicture} \Bigg)	+ 	\mathbb{P} \Bigg(
\begin{tikzpicture}[baseline=3ex,scale=1]

\node[draw=none] at (0,1) (a) {C[D]}; % 1st colum
\node[draw=none] at (0,0) (b) {D};

\node[draw=none] at (1,1) (c) {D}; % 2nd column
\node[draw=none] at (1,0) (d) {D};

\draw[->] (c) node[above] {} -- (a);
\draw[] (a) node[above] {} -- (b);
\draw[] (c) node[above] {} -- (d);

\end{tikzpicture} \Bigg)\]

\[ = \frac{1}{2} \cdot W_{d \to d} \cdot q_c \cdot q_{d|c} \cdot \frac{X^c_d}{q_c} \cdot (\alpha \cdot p)  \cdot \Bigg(  
\frac{X^d_c}{q_d} \cdot (1-p + (1-\alpha) \cdot p)  
+ \frac{X^d_d}{q_d}
\Bigg). \]

\vspace{20pt}

\[ \mathbb{P} \Bigg(						% ----- 8th term -----
\begin{tikzpicture}[baseline=3ex,scale=1]

\node[draw=none] at (0,1) (a) {C[D]}; % 1st colum
\node[draw=none] at (0,0) (b) {D};

\node[draw=none] at (1.3,1) (c) {C[D]}; % 2nd column
\node[draw=none] at (1.3,0) (d) {D};

\draw[->] (c) node[above] {} -- (a);
\draw[] (a) node[above] {} -- (b);
\draw[] (c) node[above] {} -- (d);

\end{tikzpicture} \Bigg)
= \frac{1}{2} \cdot W_{d \to d} \cdot q_c \cdot q_{c|c} \cdot \Bigg(  
\frac{X^c_d}{q_c} \cdot \frac{X^c_d}{q_c} \cdot (\alpha \cdot p)  \cdot (\alpha \cdot p)  
\Bigg) \]

\vspace{10pt}

We write the differential equation for the change in fraction of $C-C$ links on the top layer as a sum,

$$\dot{q}_{cc} = \frac{2}{k_T} \sum_E \mathbb{P}(E) \mathbb{E}(E),$$

where the summation is over all events $E$ that results in a change in $q_{c}$, and $k_T$ is the average degree of individual in the top network layer. All of the $\mathbb{P}(E)$ have been stated above. The values of $\mathbb{E} (E)$, the expected change in fraction of $C-C$ links are as follows.

\[ \mathbb{E} \Bigg(
\begin{tikzpicture}[baseline=3ex,scale=1]			% --- expected value for Wcd ---

\node[draw=none] at (0,1) (a) {D}; % 1st colum
\node[draw=none] at (0,0) (b) {};

\node[draw=none] at (1,1) (c) {C}; % 2nd column
\node[draw=none] at (1,0) (d) {};

\draw[->] (c) node[above] {} -- (a);
\draw[] (a) node[above] {} -- (b);
\draw[] (c) node[above] {} -- (d);

\end{tikzpicture} \Bigg) = 		(k_T - 1) \cdot q_{c|d} + 1,  		
\hspace{15pt} 
\mathbb{E} \Bigg(
\begin{tikzpicture}[baseline=3ex,scale=1]

\node[draw=none] at (0,1) (a) {D}; % 1st colum
\node[draw=none] at (0,0) (b) {};

\node[draw=none] at (1.3,1) (c) {D[C]}; % 2nd column
\node[draw=none] at (1.3,0) (d) {};

\draw[->] (c) node[above] {} -- (a);
\draw[] (a) node[above] {} -- (b);
\draw[] (c) node[above] {} -- (d);

\end{tikzpicture} \Bigg) = 		 (k_T - 1) \cdot q_{c|d}. \]

\[ \mathbb{E} \Bigg(
\begin{tikzpicture}[baseline=3ex,scale=1]			% --- expected value for Wcc---

\node[draw=none] at (0,1) (a) {D[C]}; % 1st colum
\node[draw=none] at (0,0) (b) {C};

\node[draw=none] at (1,1) (c) {C}; % 2nd column
\node[draw=none] at (1,0) (d) {};

\draw[->] (c) node[above] {} -- (a);
\draw[] (a) node[above] {} -- (b);
\draw[] (c) node[above] {} -- (d);

\end{tikzpicture} \Bigg) = (k_T - 1) \cdot q_{c|d} + 1,
\hspace{15pt} 
\mathbb{E} \Bigg(
\begin{tikzpicture}[baseline=3ex,scale=1]			% --- expected value for Wcc---

\node[draw=none] at (0,1) (a) {D[C]}; % 1st colum
\node[draw=none] at (0,0) (b) {C};

\node[draw=none] at (1.3,1) (c) {D[C]}; % 2nd column
\node[draw=none] at (1.3,0) (d) {C};

\draw[->] (c) node[above] {} -- (a);
\draw[] (a) node[above] {} -- (b);
\draw[] (c) node[above] {} -- (d);

\end{tikzpicture} \Bigg) = (k_T - 1) \cdot q_{c|d}. \]

\[ \mathbb{E} \Bigg(
\begin{tikzpicture}[baseline=3ex,scale=1]			% --- expected value for Wdc---

\node[draw=none] at (0,1) (a) {C}; % 1st colum
\node[draw=none] at (0,0) (b) {};

\node[draw=none] at (1,1) (c) {D}; % 2nd column
\node[draw=none] at (1,0) (d) {};

\draw[->] (c) node[above] {} -- (a);
\draw[] (a) node[above] {} -- (b);
\draw[] (c) node[above] {} -- (d);

\end{tikzpicture} \Bigg) = 		- (k_T - 1) \cdot q_{c|c},	  \hspace{15pt} \mathbb{E} \Bigg(
\begin{tikzpicture}[baseline=3ex,scale=1]

\node[draw=none] at (0,1) (a) {C}; % 1st colum
\node[draw=none] at (0,0) (b) {};

\node[draw=none] at (1.3,1) (c) {C[D]}; % 2nd column
\node[draw=none] at (1.3,0) (d) {};

\draw[->] (c) node[above] {} -- (a);
\draw[] (a) node[above] {} -- (b);
\draw[] (c) node[above] {} -- (d);

\end{tikzpicture} \Bigg) =  - (k_T - 1) \cdot q_{c|c} - 1. \]

\[ \mathbb{E} \Bigg(
\begin{tikzpicture}[baseline=3ex,scale=1]			% --- expected value for Wdd---

\node[draw=none] at (0,1) (a) {C[D]}; % 1st colum
\node[draw=none] at (0,0) (b) {D};

\node[draw=none] at (1,1) (c) {D}; % 2nd column
\node[draw=none] at (1,0) (d) {};

\draw[->] (c) node[above] {} -- (a);
\draw[] (a) node[above] {} -- (b);
\draw[] (c) node[above] {} -- (d);

\end{tikzpicture} \Bigg) = 		- (k_T - 1) \cdot q_{c|c},	  \hspace{15pt} \mathbb{E} \Bigg(
\begin{tikzpicture}[baseline=3ex,scale=1]

\node[draw=none] at (0,1) (a) {C[D]}; % 1st colum
\node[draw=none] at (0,0) (b) {D};

\node[draw=none] at (1.3,1) (c) {C[D]}; % 2nd column
\node[draw=none] at (1.3,0) (d) {D};

\draw[->] (c) node[above] {} -- (a);
\draw[] (a) node[above] {} -- (b);
\draw[] (c) node[above] {} -- (d);

\end{tikzpicture} \Bigg) =  - (k_T - 1) \cdot q_{c|c} - 1. \]

The differential equation for the change in fraction of $C$ in the bottom layer is,

			% ----- 	change in p_c 	-----

\[ \dot{p}_c = \sum_{x,y,z,u,v,w} \Bigg[ \mathbb{P} \Bigg(
\begin{tikzpicture}[baseline=3ex,scale=1]

\node[draw=none] at (0,1) (a) {}; % 1st colum
\node[draw=none] at (0,0) (b) {D};

\node[draw=none] at (1,1) (c) {}; % 2nd column
\node[draw=none] at (1,0) (d) {C};

\draw[->] (d) node[above] {} -- (b);
\draw[] (a) node[above] {} -- (b);
\draw[] (c) node[above] {} -- (d);

\end{tikzpicture} \Bigg)	+ 	\mathbb{P} \Bigg(
\begin{tikzpicture}[baseline=3ex,scale=1]

\node[draw=none] at (0,1) (a) {}; % 1st colum
\node[draw=none] at (0,0) (b) {D};

\node[draw=none] at (1,1) (c) {C}; % 2nd column
\node[draw=none] at (1,0) (d) {D[C]};

\draw[->] (d) node[above] {} -- (b);
\draw[] (a) node[above] {} -- (b);
\draw[] (c) node[above] {} -- (d);

\end{tikzpicture} \Bigg)	+ 	\mathbb{P} \Bigg(
\begin{tikzpicture}[baseline=3ex,scale=1]

\node[draw=none] at (0,1) (a) {C}; % 1st colum
\node[draw=none] at (0,0) (b) {D[C]};

\node[draw=none] at (1,1) (c) {}; % 2nd column
\node[draw=none] at (1,0) (d) {C};

\draw[->] (d) node[above] {} -- (b);
\draw[] (a) node[above] {} -- (b);
\draw[] (c) node[above] {} -- (d);

\end{tikzpicture} \Bigg)	+ 	\mathbb{P} \Bigg(
\begin{tikzpicture}[baseline=3ex,scale=1]

\node[draw=none] at (0,1) (a) {C}; % 1st colum
\node[draw=none] at (0,0) (b) {D[C]};

\node[draw=none] at (1.3,1) (c) {C}; % 2nd column
\node[draw=none] at (1.3,0) (d) {D[C]};

\draw[->] (d) node[above] {} -- (b);
\draw[] (a) node[above] {} -- (b);
\draw[] (c) node[above] {} -- (d);

\end{tikzpicture} \Bigg)      \]

			% ----- 	2nd line 	-----

\[	\hspace{15pt}	- 	 \mathbb{P} \Bigg(
\begin{tikzpicture}[baseline=3ex,scale=1]

\node[draw=none] at (0,1) (a) {}; % 1st colum
\node[draw=none] at (0,0) (b) {C};

\node[draw=none] at (1,1) (c) {}; % 2nd column
\node[draw=none] at (1,0) (d) {D};

\draw[->] (d) node[above] {} -- (b);
\draw[] (a) node[above] {} -- (b);
\draw[] (c) node[above] {} -- (d);

\end{tikzpicture} \Bigg)	- 	\mathbb{P} \Bigg(
\begin{tikzpicture}[baseline=3ex,scale=1]

\node[draw=none] at (0,1) (a) {}; % 1st colum
\node[draw=none] at (0,0) (b) {C};

\node[draw=none] at (1,1) (c) {D}; % 2nd column
\node[draw=none] at (1,0) (d) {C[D]};

\draw[->] (d) node[above] {} -- (b);
\draw[] (a) node[above] {} -- (b);
\draw[] (c) node[above] {} -- (d);

\end{tikzpicture} \Bigg)	- 	\mathbb{P} \Bigg(
\begin{tikzpicture}[baseline=3ex,scale=1]

\node[draw=none] at (0,1) (a) {D}; % 1st colum
\node[draw=none] at (0,0) (b) {C[D]};

\node[draw=none] at (1,1) (c) {}; % 2nd column
\node[draw=none] at (1,0) (d) {D};

\draw[->] (d) node[above] {} -- (b);
\draw[] (a) node[above] {} -- (b);
\draw[] (c) node[above] {} -- (d);

\end{tikzpicture} \Bigg)	- 	\mathbb{P} \Bigg(
\begin{tikzpicture}[baseline=3ex,scale=1]

\node[draw=none] at (0,1) (a) {D}; % 1st colum
\node[draw=none] at (0,0) (b) {C[D]};

\node[draw=none] at (1.3,1) (c) {D}; % 2nd column
\node[draw=none] at (1.3,0) (d) {C[D]};

\draw[->] (d) node[above] {} -- (b);
\draw[] (a) node[above] {} -- (b);
\draw[] (c) node[above] {} -- (d);

\end{tikzpicture} \Bigg) \Bigg],     \]

where the summation is over all possible strategies of the neighbours $x,y,z$ of the focal individual, and over all possible strategies $u,v,w$ of the neighbours of the non-focal individual. Detailed equations for each of the terms in the sum are as follows.

			% ---- details for terms in dot p_c ----

\[ \mathbb{P} \Bigg(						% ----- 1st term -----
\begin{tikzpicture}[baseline=3ex,scale=1]

\node[draw=none] at (0,1) (a) {}; % 1st colum
\node[draw=none] at (0,0) (b) {D};

\node[draw=none] at (1,1) (c) {}; % 2nd column
\node[draw=none] at (1,0) (d) {C};

\draw[->] (d) node[above] {} -- (b);
\draw[] (a) node[above] {} -- (b);
\draw[] (c) node[above] {} -- (d);

\end{tikzpicture} \Bigg)		=	 	 \mathbb{P} \Bigg(
\begin{tikzpicture}[baseline=3ex,scale=1]

\node[draw=none] at (0,1) (a) {C}; % 1st colum
\node[draw=none] at (0,0) (b) {D};

\node[draw=none] at (1,1) (c) {C}; % 2nd column
\node[draw=none] at (1,0) (d) {C};

\draw[->] (d) node[above] {} -- (b);
\draw[] (a) node[above] {} -- (b);
\draw[] (c) node[above] {} -- (d);

\end{tikzpicture} \Bigg)	+ 	\mathbb{P} \Bigg(
\begin{tikzpicture}[baseline=3ex,scale=1]

\node[draw=none] at (0,1) (a) {C}; % 1st colum
\node[draw=none] at (0,0) (b) {D};

\node[draw=none] at (1,1) (c) {D}; % 2nd column
\node[draw=none] at (1,0) (d) {C};

\draw[->] (d) node[above] {} -- (b);
\draw[] (a) node[above] {} -- (b);
\draw[] (c) node[above] {} -- (d);

\end{tikzpicture} \Bigg)	+ 	\mathbb{P} \Bigg(
\begin{tikzpicture}[baseline=3ex,scale=1]

\node[draw=none] at (0,1) (a) {D}; % 1st colum
\node[draw=none] at (0,0) (b) {D};

\node[draw=none] at (1,1) (c) {C}; % 2nd column
\node[draw=none] at (1,0) (d) {C};

\draw[->] (d) node[above] {} -- (b);
\draw[] (a) node[above] {} -- (b);
\draw[] (c) node[above] {} -- (d);

\end{tikzpicture} \Bigg)	+ 	\mathbb{P} \Bigg(
\begin{tikzpicture}[baseline=3ex,scale=1]

\node[draw=none] at (0,1) (a) {D}; % 1st colum
\node[draw=none] at (0,0) (b) {D};

\node[draw=none] at (1,1) (c) {D}; % 2nd column
\node[draw=none] at (1,0) (d) {C};

\draw[->] (d) node[above] {} -- (b);
\draw[] (a) node[above] {} -- (b);
\draw[] (c) node[above] {} -- (d);

\end{tikzpicture} \Bigg)\]

\[ = \frac{1}{2} \cdot \phi_{c \to d} \cdot p_d \cdot p_{c|d} \cdot \Bigg(  
\frac{X^c_d}{p_d} \cdot \frac{X^c_c}{p_c} \cdot (1-p + \alpha \cdot p) 
+ \frac{X^c_d}{p_d} \cdot \frac{X^d_c}{p_c} \cdot (1-p + \alpha \cdot p) \cdot (1-p + \alpha \cdot p) \]

\[
+ \frac{X^d_d}{p_d} \cdot \frac{X^c_c}{p_c}
+ \frac{X^d_d}{p_d} \cdot \frac{X^d_c}{p_c} \cdot (1-p + \alpha \cdot p) 
\Bigg). \]

\vspace{20pt}

\[ \mathbb{P} \Bigg(						% ----- 2nd term -----
\begin{tikzpicture}[baseline=3ex,scale=1]

\node[draw=none] at (0,1) (a) {}; % 1st colum
\node[draw=none] at (0,0) (b) {D};

\node[draw=none] at (1,1) (c) {C}; % 2nd column
\node[draw=none] at (1,0) (d) {D[C]};

\draw[->] (d) node[above] {} -- (b);
\draw[] (a) node[above] {} -- (b);
\draw[] (c) node[above] {} -- (d);

\end{tikzpicture} \Bigg)		=	 	 \mathbb{P} \Bigg(
\begin{tikzpicture}[baseline=3ex,scale=1]

\node[draw=none] at (0,1) (a) {C}; % 1st colum
\node[draw=none] at (0,0) (b) {D};

\node[draw=none] at (1,1) (c) {C}; % 2nd column
\node[draw=none] at (1,0) (d) {D[C]};

\draw[->] (d) node[above] {} -- (b);
\draw[] (a) node[above] {} -- (b);
\draw[] (c) node[above] {} -- (d);

\end{tikzpicture} \Bigg)	+ 	\mathbb{P} \Bigg(
\begin{tikzpicture}[baseline=3ex,scale=1]

\node[draw=none] at (0,1) (a) {D}; % 1st colum
\node[draw=none] at (0,0) (b) {D};

\node[draw=none] at (1,1) (c) {C}; % 2nd column
\node[draw=none] at (1,0) (d) {D[C]};

\draw[->] (d) node[above] {} -- (b);
\draw[] (a) node[above] {} -- (b);
\draw[] (c) node[above] {} -- (d);

\end{tikzpicture} \Bigg)\]

\[ = \frac{1}{2} \cdot \phi_{c \to d} \cdot p_d \cdot p_{d|d} \cdot \frac{X^c_d}{p_d} \cdot ((1-\alpha) \cdot p) \cdot \Bigg(  
\frac{X^c_d}{p_d} \cdot (1-p + \alpha \cdot p)
+ \frac{X^d_d}{p_d}
\Bigg). \]

\vspace{20pt}

\[ \mathbb{P} \Bigg(						% ----- 3rd term -----
\begin{tikzpicture}[baseline=3ex,scale=1]

\node[draw=none] at (0,1) (a) {C}; % 1st colum
\node[draw=none] at (0,0) (b) {D[C]};

\node[draw=none] at (1,1) (c) {}; % 2nd column
\node[draw=none] at (1,0) (d) {C};

\draw[->] (d) node[above] {} -- (b);
\draw[] (a) node[above] {} -- (b);
\draw[] (c) node[above] {} -- (d);

\end{tikzpicture} \Bigg)		=	 	 \mathbb{P} \Bigg(
\begin{tikzpicture}[baseline=3ex,scale=1]

\node[draw=none] at (0,1) (a) {C}; % 1st colum
\node[draw=none] at (0,0) (b) {D[C]};

\node[draw=none] at (1,1) (c) {D}; % 2nd column
\node[draw=none] at (1,0) (d) {C};

\draw[->] (d) node[above] {} -- (b);
\draw[] (a) node[above] {} -- (b);
\draw[] (c) node[above] {} -- (d);

\end{tikzpicture} \Bigg)	+ 	\mathbb{P} \Bigg(
\begin{tikzpicture}[baseline=3ex,scale=1]

\node[draw=none] at (0,1) (a) {C}; % 1st colum
\node[draw=none] at (0,0) (b) {D[C]};

\node[draw=none] at (1,1) (c) {C}; % 2nd column
\node[draw=none] at (1,0) (d) {C};

\draw[->] (d) node[above] {} -- (b);
\draw[] (a) node[above] {} -- (b);
\draw[] (c) node[above] {} -- (d);

\end{tikzpicture} \Bigg)\]

\[ = \frac{1}{2} \cdot \phi_{c \to c} \cdot p_d \cdot p_{c|d} \cdot \frac{X^c_d}{p_d} \cdot ((1-\alpha) \cdot p) \cdot \Bigg(  
\frac{X^d_c}{p_c} \cdot (1-p + \alpha \cdot p)  
+ \frac{X^c_c}{p_c} 
\Bigg). \]

\vspace{20pt}

\[ \mathbb{P} \Bigg(						% ----- 4th term -----
\begin{tikzpicture}[baseline=3ex,scale=1]

\node[draw=none] at (0,1) (a) {C}; % 1st colum
\node[draw=none] at (0,0) (b) {D[C]};

\node[draw=none] at (1.3,1) (c) {C}; % 2nd column
\node[draw=none] at (1.3,0) (d) {D[C]};

\draw[->] (d) node[above] {} -- (b);
\draw[] (a) node[above] {} -- (b);
\draw[] (c) node[above] {} -- (d);

\end{tikzpicture} \Bigg)
= \frac{1}{2} \cdot \phi_{c \to c} \cdot p_d \cdot p_{d|d} \cdot \Bigg(  
\frac{X^c_d}{p_d} \cdot \frac{X^c_d}{p_d} \cdot ((1-\alpha) \cdot p)  \cdot ((1-\alpha) \cdot p)  
\Bigg). \]

\vspace{20pt}

\[ \mathbb{P} \Bigg(						% ----- 5th term -----
\begin{tikzpicture}[baseline=3ex,scale=1]

\node[draw=none] at (0,1) (a) {}; % 1st colum
\node[draw=none] at (0,0) (b) {C};

\node[draw=none] at (1,1) (c) {}; % 2nd column
\node[draw=none] at (1,0) (d) {D};

\draw[->] (d) node[above] {} -- (b);
\draw[] (a) node[above] {} -- (b);
\draw[] (c) node[above] {} -- (d);

\end{tikzpicture} \Bigg)		=	 	 \mathbb{P} \Bigg(
\begin{tikzpicture}[baseline=3ex,scale=1]

\node[draw=none] at (0,1) (a) {C}; % 1st colum
\node[draw=none] at (0,0) (b) {C};

\node[draw=none] at (1,1) (c) {C}; % 2nd column
\node[draw=none] at (1,0) (d) {D};

\draw[->] (d) node[above] {} -- (b);
\draw[] (a) node[above] {} -- (b);
\draw[] (c) node[above] {} -- (d);

\end{tikzpicture} \Bigg)	+ 	\mathbb{P} \Bigg(
\begin{tikzpicture}[baseline=3ex,scale=1]

\node[draw=none] at (0,1) (a) {C}; % 1st colum
\node[draw=none] at (0,0) (b) {C};

\node[draw=none] at (1,1) (c) {D}; % 2nd column
\node[draw=none] at (1,0) (d) {D};

\draw[->] (d) node[above] {} -- (b);
\draw[] (a) node[above] {} -- (b);
\draw[] (c) node[above] {} -- (d);

\end{tikzpicture} \Bigg)	+ 	\mathbb{P} \Bigg(
\begin{tikzpicture}[baseline=3ex,scale=1]

\node[draw=none] at (0,1) (a) {D}; % 1st colum
\node[draw=none] at (0,0) (b) {C};

\node[draw=none] at (1,1) (c) {C}; % 2nd column
\node[draw=none] at (1,0) (d) {D};

\draw[->] (d) node[above] {} -- (b);
\draw[] (a) node[above] {} -- (b);
\draw[] (c) node[above] {} -- (d);

\end{tikzpicture} \Bigg)	+ 	\mathbb{P} \Bigg(
\begin{tikzpicture}[baseline=3ex,scale=1]

\node[draw=none] at (0,1) (a) {D}; % 1st colum
\node[draw=none] at (0,0) (b) {C};

\node[draw=none] at (1,1) (c) {D}; % 2nd column
\node[draw=none] at (1,0) (d) {D};

\draw[->] (d) node[above] {} -- (b);
\draw[] (a) node[above] {} -- (b);
\draw[] (c) node[above] {} -- (d);

\end{tikzpicture} \Bigg)\]

\[ = \frac{1}{2} \cdot \phi_{d \to c} \cdot p_c \cdot p_{d|c} \cdot \Bigg(  
\frac{X^c_c}{p_c} \cdot \frac{X^c_d}{p_d} \cdot (1-p + \alpha \cdot p) 
+ \frac{X^c_c}{p_c} \cdot \frac{X^d_d}{p_d} \]

\[
+ \frac{X^d_c}{p_c} \cdot \frac{X^c_d}{p_d} \cdot (1-p + \alpha \cdot p) \cdot (1-p + \alpha \cdot p) 
+ \frac{X^d_c}{p_c} \cdot \frac{X^d_d}{p_d} \cdot (1-p + \alpha \cdot p) 
\Bigg). \]

\[ \mathbb{P} \Bigg(						% ----- 6th term -----
\begin{tikzpicture}[baseline=3ex,scale=1]

\node[draw=none] at (0,1) (a) {}; % 1st colum
\node[draw=none] at (0,0) (b) {C};

\node[draw=none] at (1,1) (c) {D}; % 2nd column
\node[draw=none] at (1,0) (d) {C[D]};

\draw[->] (d) node[above] {} -- (b);
\draw[] (a) node[above] {} -- (b);
\draw[] (c) node[above] {} -- (d);

\end{tikzpicture} \Bigg)		=	 	 \mathbb{P} \Bigg(
\begin{tikzpicture}[baseline=3ex,scale=1]

\node[draw=none] at (0,1) (a) {D}; % 1st colum
\node[draw=none] at (0,0) (b) {C};

\node[draw=none] at (1,1) (c) {D}; % 2nd column
\node[draw=none] at (1,0) (d) {C[D]};

\draw[->] (d) node[above] {} -- (b);
\draw[] (a) node[above] {} -- (b);
\draw[] (c) node[above] {} -- (d);

\end{tikzpicture} \Bigg)	+ 	\mathbb{P} \Bigg(
\begin{tikzpicture}[baseline=3ex,scale=1]

\node[draw=none] at (0,1) (a) {C}; % 1st colum
\node[draw=none] at (0,0) (b) {C};

\node[draw=none] at (1,1) (c) {D}; % 2nd column
\node[draw=none] at (1,0) (d) {C[D]};

\draw[->] (d) node[above] {} -- (b);
\draw[] (a) node[above] {} -- (b);
\draw[] (c) node[above] {} -- (d);

\end{tikzpicture} \Bigg)\]

\[ = \frac{1}{2} \cdot \phi_{d \to c} \cdot p_c \cdot p_{c|c} \cdot \frac{X^d_c}{p_c} \cdot ((1-\alpha) \cdot p) \cdot \Bigg(  
\frac{X^d_c}{p_c} \cdot (1-p + \alpha \cdot p) 
+ \frac{X^c_c}{p_c} 
\Bigg). \]

\vspace{20pt}

\[ \mathbb{P} \Bigg(						% ----- 7th term -----
\begin{tikzpicture}[baseline=3ex,scale=1]

\node[draw=none] at (0,1) (a) {D}; % 1st colum
\node[draw=none] at (0,0) (b) {C[D]};

\node[draw=none] at (1,1) (c) {}; % 2nd column
\node[draw=none] at (1,0) (d) {D};

\draw[->] (d) node[above] {} -- (b);
\draw[] (a) node[above] {} -- (b);
\draw[] (c) node[above] {} -- (d);

\end{tikzpicture} \Bigg)		=	 	 \mathbb{P} \Bigg(
\begin{tikzpicture}[baseline=3ex,scale=1]

\node[draw=none] at (0,1) (a) {D}; % 1st colum
\node[draw=none] at (0,0) (b) {C[D]};

\node[draw=none] at (1,1) (c) {C}; % 2nd column
\node[draw=none] at (1,0) (d) {D};

\draw[->] (d) node[above] {} -- (b);
\draw[] (a) node[above] {} -- (b);
\draw[] (c) node[above] {} -- (d);

\end{tikzpicture} \Bigg)	+ 	\mathbb{P} \Bigg(
\begin{tikzpicture}[baseline=3ex,scale=1]

\node[draw=none] at (0,1) (a) {D}; % 1st colum
\node[draw=none] at (0,0) (b) {C[D]};

\node[draw=none] at (1,1) (c) {D}; % 2nd column
\node[draw=none] at (1,0) (d) {D};

\draw[->] (d) node[above] {} -- (b);
\draw[] (a) node[above] {} -- (b);
\draw[] (c) node[above] {} -- (d);

\end{tikzpicture} \Bigg)\]

\[ = \frac{1}{2} \cdot \phi_{d \to d} \cdot p_c \cdot p_{d|c} \cdot \frac{X^d_c}{p_c} \cdot ((1-\alpha) \cdot p) \cdot \Bigg(  
\frac{X^c_d}{p_d} \cdot (1-p + \alpha \cdot p) 
+ \frac{X^d_d}{p_d} 
\Bigg). \]

\vspace{20pt}

\[ \mathbb{P} \Bigg(						% ----- 8th term -----
\begin{tikzpicture}[baseline=3ex,scale=1]

\node[draw=none] at (0,1) (a) {D}; % 1st colum
\node[draw=none] at (0,0) (b) {C[D]};

\node[draw=none] at (1.3,1) (c) {D}; % 2nd column
\node[draw=none] at (1.3,0) (d) {C[D]};

\draw[->] (d) node[above] {} -- (b);
\draw[] (a) node[above] {} -- (b);
\draw[] (c) node[above] {} -- (d);

\end{tikzpicture} \Bigg)
= \frac{1}{2} \cdot \phi_{d \to d} \cdot p_c \cdot p_{c|c} \cdot \Bigg(  
\frac{X^d_c}{p_c} \cdot \frac{X^d_c}{p_c} \cdot ((1-\alpha) \cdot p)  \cdot ((1-\alpha) \cdot p)  
\Bigg). \]

\vspace{10pt}

We write the differential equation for the change in fraction of $C-C$ links on the bottom layer as a sum,

$$\dot{p}_{cc} = \frac{2}{k_B} \sum_E \sum_{x,y,z,u,v,w} \mathbb{P}(E) \mathbb{E}(E),$$

where the summation is over all possible strategies of the neighbours $x,y,z$ of the focal individual, over all possible strategies $u,v,w$ of the neighbours of the non-focal individual, and all events $E$ that results in a change in $p_{c}$. $k_B$ is the average degree of the bottom layer. All of the $\mathbb{P}(E)$ have been defined above. Let $n(x,y,z)$ be the number of $C$ among neighbour strategies $x,y,z$. Then $\mathbb{E} (E)$, the expected change in fraction of $C-C$ links for each event $E$ are as follows.

\[ \mathbb{E} \Bigg(
 \Bigg) =  - n(x,y,z) - 1. \]

The differential equation for the change in fraction of individuals with strategy $C$ on the top layer and $C$ in the bottom layer is,

			% ----- 	change in cXc	-----

\[ \dot{X}^c_c = - \mathbb{P} \Bigg(
%
 \Bigg). \]

Details for each term in the sum are as follows.

\[ \mathbb{P} \Bigg(						% ----- cXc 1st term -----
%
 \Bigg)\]

\[ = \frac{1}{2} \cdot \phi_{c \to c} \cdot p_d \cdot p_{c|d} \cdot \frac{X^c_d}{p_d} \cdot ((1-\alpha) \cdot p)   \cdot \Bigg(  
\frac{X^d_c}{p_c} \cdot (1-p + \alpha \cdot p)    
+ \frac{X^c_c}{p_c}
\Bigg). \]

The differential equation for the change in fraction of individuals with strategy $D$ on the top layer and $C$ in the bottom layer is,

			% ----- 	change in dXc	-----

\[ \dot{X}^d_c = - \mathbb{P} \Bigg(
%
 \Bigg).      \]

Details for each term in the sum are as follows.

\[ \mathbb{P} \Bigg(						% ----- dXc 1st term -----
%
 \Bigg)
 = \frac{1}{2} \cdot \phi_{c \to d} \cdot p_d \cdot p_{d|d} \cdot \frac{X^d_d}{p_d} \cdot
\frac{X^c_d}{p_d} \cdot ((1-\alpha) \cdot p). \]

\end{document}